\documentstyle[12pt,epsf,graphicx,rotating,epsfig]{article}
\newcommand{\be}{\begin{eqnarray}}

\newcommand{\ee}{\end{eqnarray}}

\makeindex

\begin{document}

\title{
RHIC Physics: The Quark Gluon Plasma and The Color
Glass Condensate: 4 Lectures\footnote{Lectures Delivered at the BARC
Workshop ``Mesons and Quarks'',  Mumbai, India; Jan.-Feb. 2003}
}
\author{Larry McLerran\\
{\small\it Physics Department 
PO Box 5000 Brookhaven National Laboratory  
Upton, NY 11973 USA }\\
}
\maketitle

\begin{abstract} The purpose of these lectures is to provide an
introduction to the physics issues which are being studied in
the RHIC heavy ion program.  These  center around
the production of new states of matter.  The Quark Gluon Plasma
is thermal matter which once existed in the big bang which may be made
at RHIC.  The Color Glass Condensate is a universal form of matter
which controls the high energy limit of strong interactions.
Both such forms of matter might be produced and probed at RHIC.
\end{abstract}

\section{Introduction}

These lectures will introduce the listener to the physics issues behind the
experimental heavy ion program at RHIC.  This program involves the collisions
of protons on protons, deuterons on nuclei, and nuclei on nuclei.  
The collision
energy is of order 200 GeV per nucleon in the center of mass.   The goal of 
these experimental studies is to produce new forms of matter.  This may be
a Quark Gluon Plasma or a Color Glass Condensate. The properties of these
forms of matter are described below.

The outline of these lectures is
\begin{itemize}
\item{\Large\bf New States of Matter}\\
In the first lecture I describe the new forms of matter which 
may be produced in heavy ion collisions.  These are the
Quark Gluon Plasma and the Color Glass Condensate.
\vspace{0.1in}
\item{\Large\bf Space Time Dynamics}\\
This lecture describes the space-time dynamics of high energy
heavy ion collisions.  In this lecture, I illustrate how high energy density
matter might be formed.
\vspace{0.1in}
\item{\Large\bf Experiment and Theory}\\
In this lecture, I show  how various experimental measurements
might teach us about the properties of matter. 

\item{\Large\bf The Color Glass Condensate}\\
In this lecture, some aspects of the Color Glass
Condensate are developed, in particular the renormalization
group equations.

\end{itemize}

\section{Lecture I: High Density Matter}

\subsection{The Goals of RHIC}
The goal of nuclear physics has traditionally been to study matter at
densities of the order of those in the atomic nucleus,
\be
	\epsilon \sim .15 ~GeV/Fm^3
\ee
High energy nuclear physics has extended this study to energy densities 
several orders of magnitude higher. This extension includes the study of
matter inside ordinary strongly interacting particles, such as the 
proton and the neutron, and producing new forms of matter at much higher
energy densities in high energy collisions of nuclei with nuclei, and various 
other probes. 

RHIC is a multi-purpose machine which can address at least three central 
issues of high energy nuclear physics.  These are:  

\begin{itemize}

\item 
{ \bf The production
of matter at energy densities one to two orders of magnitude higher than
that of nuclear matter and the study of its properties.}\\ This matter
is at such high densities that it is only simply described in terms of
quarks and gluons and is generically referred to as the Quark Gluon Plasma.
The study of this matter may allow us to better
understand the origin of the masses of ordinary particles such as nucleons,
and of the confinement of quarks and gluons into hadrons.
The Quark Gluon Plasma will be described below.\cite{qgp}

\item
{\bf The study of the matter which controls high energy strong interactions.}\\
This matter is believed to be universal (independent of the hadron), 
and exists over sizes large compared to the typical microphysics size
scales important for high energy strong interactions.  
(The microphysics size scale
here is about $1~Fm$ and the microphysics time scale is the time
it takes light to fly $1~Fm$, $t \sim 10^{-23}~sec$.)  It is called a Color
Glass Condensate because it is composed of colored particles,  
evolves on time scales long compared to microphysics time scales and therefore
has properties similar to glasses, and a condensate since the phase space
density of gluons is very high. The study of this matter may allow us 
to better understand the typical features of strong interactions
when they are truly strong, a 
problem which has eluded a basic understanding since strong interactions
were first discovered.   
The Color Glass Condensate will be
described below.\cite{cgc}

\item
{\bf The study of the structure of the proton, most notably spin.}\\  

The structure of the proton and neutron is important as these particles
form the ordinary matter from which we are composed.  We would like to 
understand how valence quantum numbers such as baryon number, charge and spin
are distributed.  RHIC has an active program to study the spin of the 
proton.\cite{spin}

\end{itemize}

Because I was asked to provide lectures on the heavy ion program at
RHIC,  I shall discuss only
the first two issues.

\subsection{The Quark Gluon Plasma}

This section describes what is the Quark Gluon Plasma, why it is important
for astrophysics and cosmology, and why it provides a laboratory
in which one can study the origin of mass and of confinement.\cite{qgp} 
 
\subsubsection{What is the Quark Gluon Plasma?}

\begin{figure}[ht]
    \begin{center}
        \includegraphics[width=0.50\textwidth]{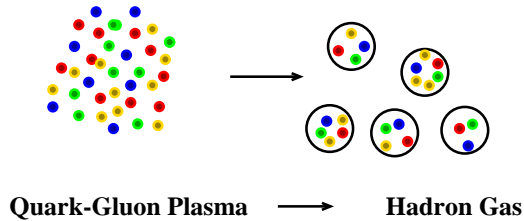}
        \caption{As the energy density is decreased,
                 the Quark Gluon Plasma condenses 
          into a low density gas of hadrons. Quarks are
red, green or blue and gluons are yellow. }\label{qgptohg}
    \end{center}
\end{figure}

Matter at low energy densities is composed of electrons, 
protons and neutrons.  If we heat the system, we might
produce thermal excitations which include light mass 
strongly interacting particles such as the pion.
Inside the protons, neutrons and other strongly interacting particles
are quarks and gluons.  If we make the matter have high enough
energy density, the protons, nucleons and other particles overlap and 
get squeezed so tightly that their constituents are free to roam the system
without being confined inside hadrons.\cite{earlywork}  
At this density, there is deconfinement and the
system is called a Quark Gluon Plasma. This is shown in Fig. \ref{qgptohg}

As the energy density gets to be very large, the interactions between the 
quarks and gluons become weak.  This is a consequence of the asymptotic
freedom of strong interactions:  At short distances the strong interactions
become weak.

\begin{figure}[ht]
    \begin{center}
        \includegraphics[width=0.80\textwidth]{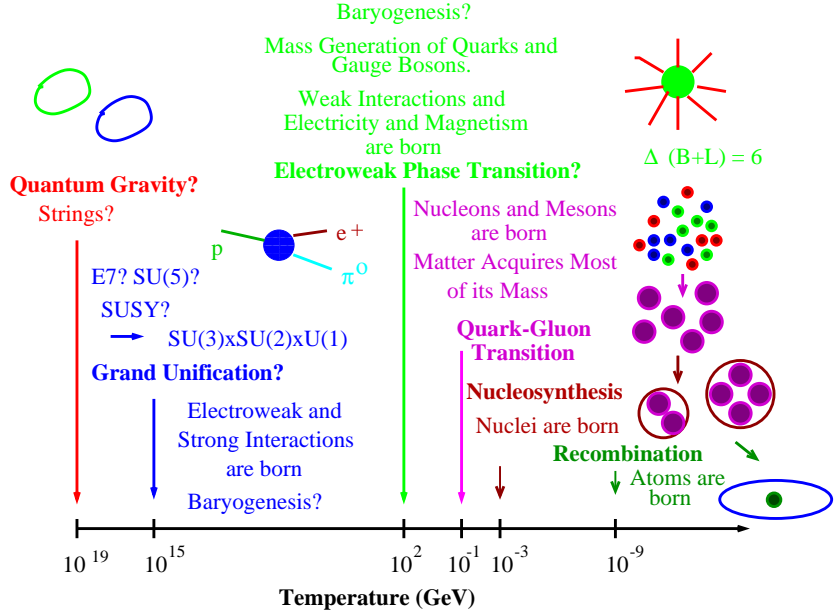}
        \caption{The various forms of matter,
and the types of physics which are probed during the big bang. }\label{bigbang}
    \end{center}
\end{figure}

The Quark Gluon Plasma surely existed during the big bang.  
In  Fig \ref{bigbang}, 
the various stages of evolution in the big bang are shown.
At the earliest times in the big bang, 
temperatures are of order $T \sim 10^{19} ~GeV$, 
quantum gravity is important, and
despite the efforts of several generations of string theorists, we
have little understanding.  At somewhat
lower temperatures, perhaps there is the grand unification of
all the forces, except gravity.  It might be possible
that the baryon number of the universe is generated at this
temperature scale.  At much lower temperatures, of order
$T \sim 100 ~GeV$, electroweak symmetry breaking takes place.
It is possible here that the baryon asymmetry of the
universe might be produced.  At temperatures of order $T \sim 1~GeV$,
quarks and gluons become confined into hadrons.  This is the temperature
range RHIC is designed to study.  At $T \sim 1 ~MeV$,
the light elements are made.  This temperature corresponds to an 
energy range which has been much studied, and is the realm of 
conventional nuclear physics.  At temperatures of the order of
an electron volt, corresponding to the binding energies of
electrons in atoms, the universe changes from an ionized gas
to a lower pressure gas of atoms, and structure begins to form.

\begin{figure}[ht]
    \begin{center}
        \includegraphics[width=0.50\textwidth]{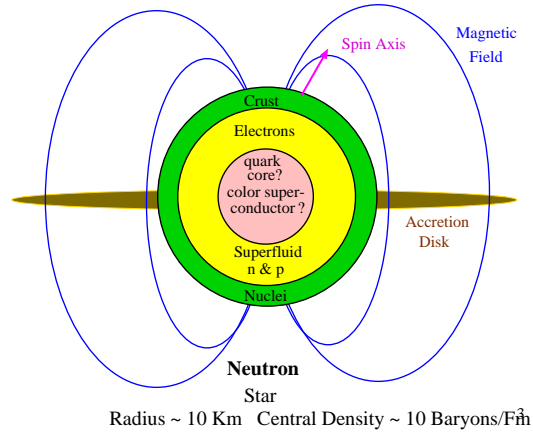}
        \caption{A spinning neutron star }
\label{neutronstar}
    \end{center}
\end{figure}

The Quark Gluon Plasma is formed at energy densities of
order $1~GeV/Fm^3$.  Matter at such energy densities probably exists
inside the cores of neutron stars as shown in Fig. \ref{neutronstar}.
Neutron stars are objects of about $10~Km$ in radius and 
are composed of extremely high energy density matter.  The typical energy
density in the core is of the order of $1~GeV/Fm^3$, and approaches zero
at the surface.  Unlike the matter in the big bang, this matter is
cold and has temperature small compared to the Fermi
energies of quarks.  It is a cold, degenerate gas of quarks.  
At lower densities, this matter converts into a cold gas of nucleons.

Hot and dense matter with energy density of order $1~GeV/Fm^3$
may have occurred in the supernova explosion which led to the neutron
star's formation.  It may also occur in collisions of neutron stars
and black holes, and may be the origin of the mysterious gamma
ray bursters.  (Gamma ray bursters are believed to be starlike objects
which convert of the order of their entire mass in to gamma rays.)

\subsubsection{The Quark Gluon Plasma and Ideal Gasses}

At very high energy temperatures, the coupling constant of QCD becomes weak.
A gas of particles should to a good approximation become an ideal
gas.  Each species of particle contributes to the energy density 
of an ideal gas as
\be
 \epsilon = \int {{d^3p} \over {(2\pi)^3}} \sum_i  {E_i \over {e^{\beta E_i}
\pm 1}}
\ee
where the $-$ is for Bosons and the $+$ for Fermions.
The energy of each particle is $E_i$.  At high temperatures, masses can
be ignored, and the factor of $\pm 1$ in the denominator turns out to make
a small difference.  One finds therefore that
\be
	\epsilon \sim {\pi^2 \over {30}} N T^4
\ee
where $N$ is the number of particle degrees of freedom.  At low 
temperatures when masses are important, only the lowest 
mass strongly interacting particle degree of freedom contributes, the
pion, and the energy density approaches zero as $\epsilon \sim e^{-m_\pi /T}$.
For an ideal gas of pions, the number of pion degrees of freedom are three.
For a quark gluon plasma there are two helicities and eight colors for
each gluon, and for each quark, three colors, 2 spins and a quark-antiquark
pair.  The number of degrees of freedom is $N \sim 2 \times 8 + 4 \times 3 
\times
N_F$ where $N_F$ is the number of important quark flavors, which is
about $3$ if the temperature is below the charm quark mass\, 
so that $N \sim 50$.

There is about an order of magnitude change in the number of 
degrees of freedom between a hadron gas and a Quark Gluon Plasma.
This is because the degrees of freedom of the QGP include color.  In the large
$N_{color}$ limit, the number of degrees of freedom of the plasma
are proportional to $N_{color}^2$, and in the confined phase is of order $1$.
In this limit, the energy density has an infinite discontinuity
at the phase transition.  There would be a limiting temperature for 
the hadronic world in the limit for which $N_{color} \rightarrow \infty$, 
since at some temperature the energy 
density would go to infinity.  This is the Hagedorn limiting temperature.
(In the real world $N_{color}$ is three, and there is 
a temperature at which the energy density changes by an order
of magnitude in a narrow range.)

\subsubsection{The Quark Gluon Plasma and Fundamental Physics Issues}

The nature of matter at high densities is an issue of fundamental interest.
Such matter occurred during the big bang, and it is the ultimate
and universal state of matter at very high energy densities.

\begin{figure}[ht]
    \begin{center}
        \includegraphics[width=0.60\textwidth]{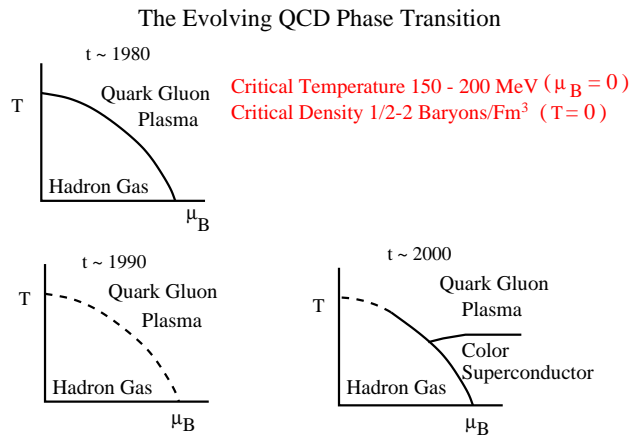}
        \caption{A phase diagram for QCD
collisions.}\label{phasediagram}
    \end{center}
\end{figure}

A hypothetical phase diagram for QCD is shown in Fig. \ref{phasediagram}.
The vertical axis is temperature, and the horizontal is a measure
of the matter or baryon number density, the baryon number
chemical potential.\cite{lgt}
The solid lines indicate a first order phase transition, and
the dashed line a rapid cross over.  It is not known 
for sure whether or not the region
marked cross over is or is not a true first order phase transition.
There are analytic arguments for the properties of matter at high density, 
but numerical computation are 
of insufficient resolution.  At high temperature and fixed baryon number
density, there are both analytic arguments and numerical computations of good
quality.
At high density and fixed
temperature, one goes into a superconducting phase, perhaps multiple
phases of superconducting quark matter.  
At high temperature and fixed baryon number density, the degrees of 
freedom are those of a Quark Gluon Plasma.

I have shown this phase diagram as a function of time.  What this means is
that at various times people thought they knew what the phase diagram was.
As time evolved, the picture changed.  The latest ideas are marked with the
date 2000.  The point of doing this is to illustrate that theoretical
ideas in the absence of experiment change with time.  Physics is essentially
an experimental science, and it is very difficult to appreciate 
the richness which nature allows without knowing from experiment what 
is possible.

Much of the information we have about QCD at finite energy density comes
from lattice gauge theory numerical simulation.\cite{lgt}  
To see how lattice gauge
theory works, recall that at finite temperature, the Grand Canonical
Ensemble is given by
\be
	Z = Tr~ e^{-\beta H}
\ee
This is similar to computing 
\be
	Z = < e^{-itH} >
\ee
where $-it = \beta$.  That is we compute the expectation value of the
time evolution operator for imaginary time.  This object has a 
path integral representation, which has been described to you in your
elementary field theory text books.  Under the change of variables,
the action becomes $iS = i \int dt L \rightarrow S = -\int_0^\beta d\tau L$.
Here $L$ is the Lagrangian. 

 The Grand Canonical Ensemble has
the representation
\be
	Z = \int [dA] e^{-S[A]}
\ee
for a system of pure gluons.  The gluon fields satisfy periodic
boundary conditions due to the trace in the definition of the Grand
Canonical Ensemble.  (Fermions may also be included, although
the path integral is more complicated, and the fermion fields are
required to satisfy antiperiodic boundary conditions.)  Expectation values
are computed as
\be
	<0> = {{Tr~O e^{-\beta H}} \over {Tr~e^{-\beta H}}}
\ee

The way that lattice Monte Carlo simulates the Grand Canonical Ensemble
is by placing all of the fields on a finite grid, so the
path integral becomes finite dimensional.  Then field configurations
are selectively sampled, as weighted by their action.  This works because the
factor of $e^{-\beta H}$ is positive and real.  (The method has essential
complications for finite density systems, since there the action
becomes complex.)

\begin{figure}[ht]
    \begin{center}
        \includegraphics[width=0.60\textwidth]{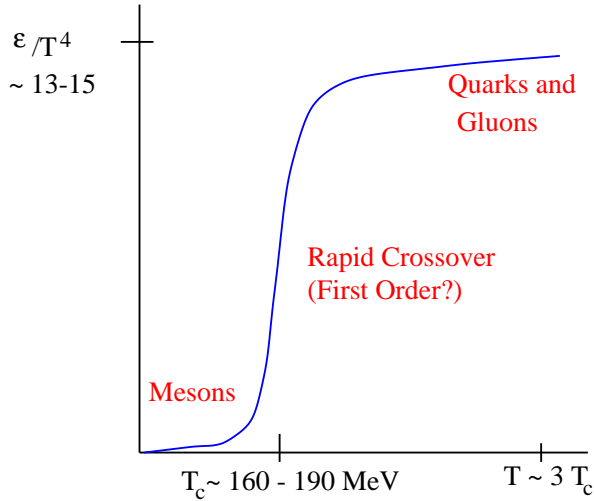}
        \caption{The energy density scaled by $T^4$
as a function of temperature.}\label{evst}
    \end{center}
\end{figure}

Lattice gauge theory numerical studies, and analytic studies have taught us 
much about the properties of these various phases of matter.\cite{lgt}  
There have been detailed computations of the energy density
as a function of temperature.  In Fig. \ref{evst}
the energy density scaled by $T^4$ is plotted.  This is
essentially the number of degrees of freedom of the system as function
of $T$.  At a temperature of $T_c \sim 160-190~MeV$ the number of degrees
of freedom changes very rapidly, possibly discontinuously.  This is the
location of the transition from the hadron gas to the quark
gluon plasma.

\begin{figure}[ht]
    \begin{center}
        \includegraphics[width=0.60\textwidth]{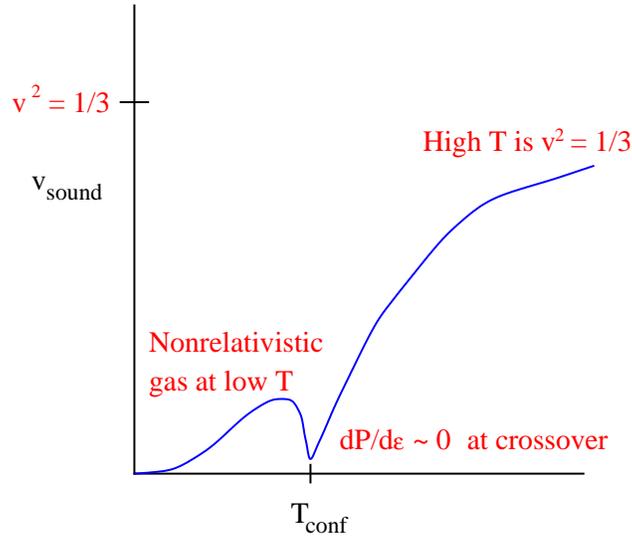}
        \caption{The sound velocity as a function of temperature.}
\label{sound}
    \end{center}
\end{figure}

In Fig. \ref{sound}, the sound velocity is plotted as
a function of temperature.  The sound velocity increases at high temperature
asymptoting to its ideal gas value of $v_{sound}^2 \sim 1/3$.  Near
the phase transition, it become very small.  This is because the
energy density jumps at the transition temperature, but the pressure
must be smooth and continuous.  The sound velocity squared is $d\epsilon/dP$.

Lattice Monte-Carlo simulation has also studied how the phase transition
is related to the confining force.  In a theory with only gluons, the 
potential for sources of fundamental representation color charge
grows linearly in the confined phase.  (With dynamical fermions, the
potential stops rising at some distance when it is energetically
favorable to produce quark-antiqaurk pairs which short out the potential.)

We can understand how confinement might disappear at high temperature.
A finite temperature, there is a symmetry of the pure gluon Yang-Mills system.
Consider a Wilson line which propagates
from $(0,\vec{x})$ to the point $(\beta,\vec{x})$  
A wilson line is a path ordered phase,
\be
       L(x) = P \exp{\int_0^\beta dt A^0(t,\vec{x}) }
\ee
One can show that
the expectation value of this line gives the free energy of an isolated
quark:
\be
	e^{-\beta F} = { 1\over N_c} < tr(L(x))>
\ee
Now consider gauge transformations which maintain
the periodic boundary conditions on the gauge fields (required by
the trace in the definition of the Grand Canonical Ensemble).  The most
general gauge transformation which does this is not periodic but solves
\be
	U(\beta, \vec{x}) = Z U(0,\vec{x})
\ee
One can show that $[z, \tau^a] = 0$, and that $\nabla^i Z = 0$.
$Z$ is an element of the gauge group so that $det Z = 1$.  These conditions
require that 
\be
	Z = e^{2 \pi ij/N_c}
\ee

This symmetry under non-periodic gauge transformations is global, that is
it does not depend upon the position in space.  It may be broken.
If it is realized, the free energy of a quark must be infinite since
$L \rightarrow ZL$ under this transformation, and $<L> = 0$.
If the symmetry is broken, quarks can be free.

\begin{figure}[ht]
    \begin{center}
        \includegraphics[width=0.60\textwidth]{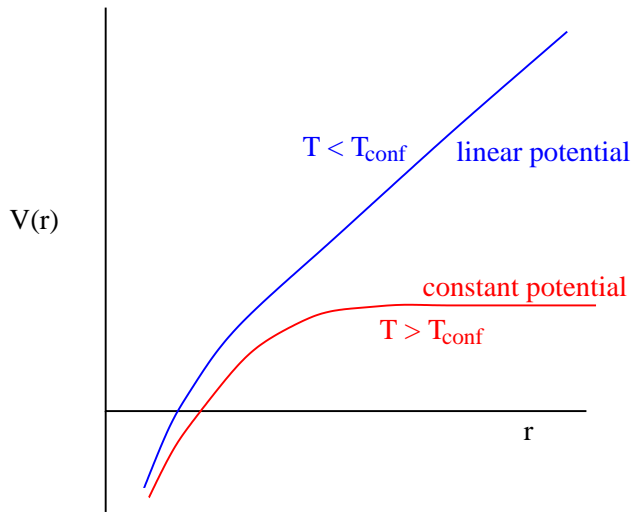}
        \caption{The potential in pure gauge
theory as a function of temperature.}\label{vr}
    \end{center}
\end{figure}

Lattice gauge computations have measured the quark-antiquark potential
as a function of $T$, and at the deconfinement temperature, the potential
changes from linear at infinity to constant.  This is shown in Fig. \ref{vr}.

\begin{figure}[ht]
    \begin{center}
        \includegraphics[width=0.60\textwidth]{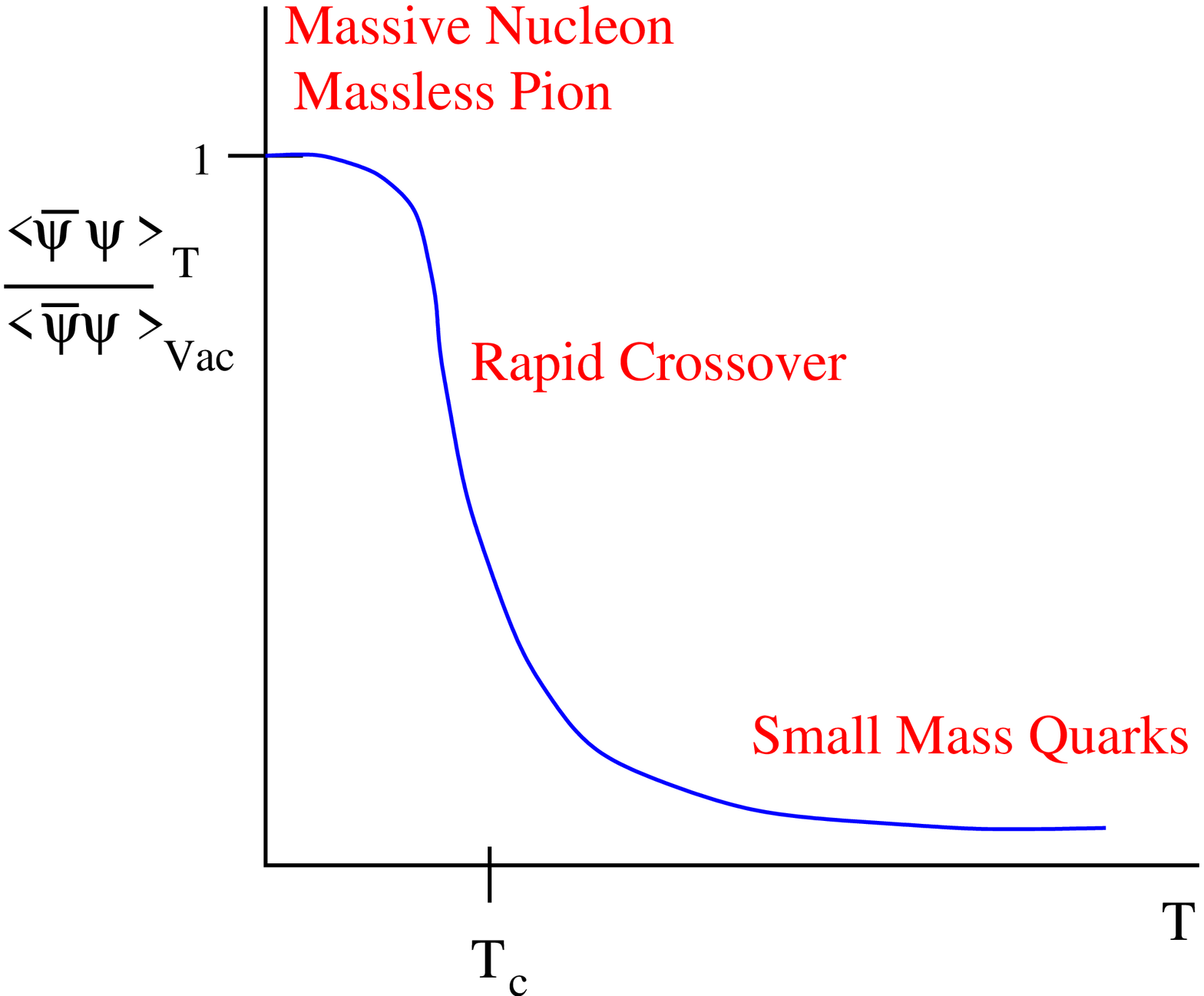}
        \caption{The chiral order parameter $<\overline \Psi \Psi >$
as a function of temperature.}\label{psivst}
    \end{center}
\end{figure}

In addition to confinement-deconfinement, there is an additional
symmetry which might become realized at high temperatures.  In nature,
the up and down quark masses are almost zero.  This leads to a chiral symmetry,
which is the rotation of fermion fields by $e^{i \gamma_5 \theta}$.
This symmetry if realized would require that either baryons are massless
or occur in parity doublets.  Neither is realized in nature.  The nucleon
has a mass of about $1~GeV$ and has no opposite parity partner of
almost equal mass.  It is believed that this symmetry becomes broken,
and as a consequence, the nucleon acquires mass, and that the pion
becomes an almost massless Goldstone boson.  It turns out that at
the confinement-deconfinement phase transition, chiral symmetry is
restored.  This is seen in Fig. \ref{psivst}, where a quantity 
proportional to the nucleon mass is plotted as a function of T.

\begin{figure}[ht]
    \begin{center}
        \includegraphics[width=0.60\textwidth]{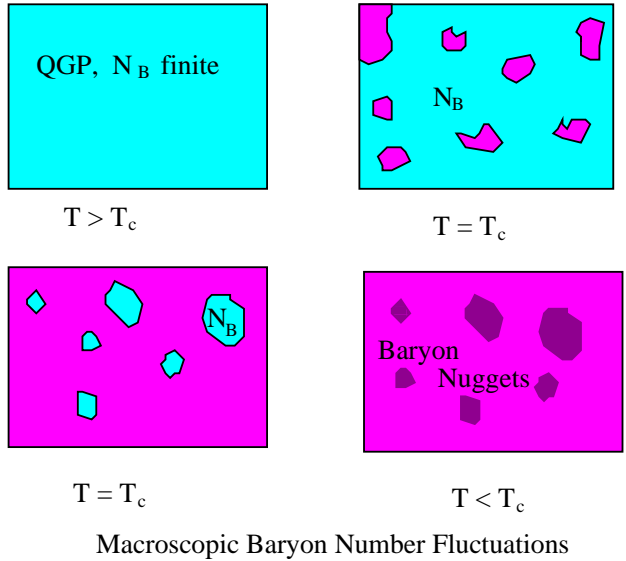}
        \caption{Formation of large scale baryon number
fluctuations at the QCD phase transition.}\label{nuggets}
    \end{center}
\end{figure}

The chiral symmetry restoration phase transition can have interesting
dynamical consequences.  In the confined phase, the mass of a nucleon is of
order $N_c \Lambda_{QCD}$, but in the deconfined phase is of order $T$.
Therefore in the confined phase, the Boltzman weight $e^{-M/T}$ is very small.
Imagine what happens as we go through the phase transition starting at a 
temperature above $T_c$.  At first the system is entirely in QGP.  As the 
system expands, a mixed phase of droplets of QGP and droplets of hadron
gas form.  The nucleons like to stay in the QGP because their Boltzman weight is
larger.  As the system expands further, the droplets of QGP shrink, but 
most of the baryon number is concentrated in them.  At the
end of the mixed phase, one has made large scale fluctuations in the baryon 
number.  This scenario is shown in Fig. \ref{nuggets}

\begin{figure}[ht]
    \begin{center}
        \includegraphics[width=0.60\textwidth]{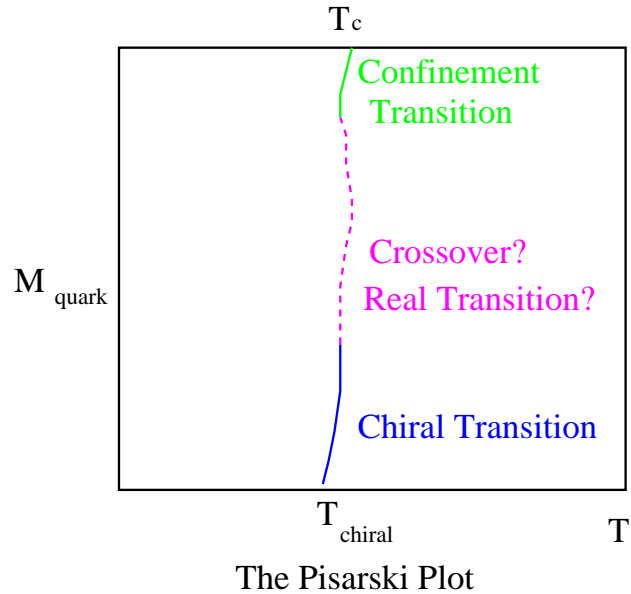}
        \caption{The phase diagram of QCD as a function
of fermion mass.}\label{pisarski}
    \end{center}
\end{figure}

The confinement-deconfinement phase transition 
and the chiral symmetry restoration phase transition might be logically
disconnected from one another.  The confinement-deconfinement phase transition
is related to a symmetry when the quark masses are infinite.  The chiral
transition is related to a symmetry when the quarks are massless.
As a function of mass, one can follow the evolution of the phase
transitions.  At large and small masses there is a real phase transition
marked by a discontinuity in physical quantities. At intermediate masses,
there is probably a rapid transition, but not a real
phase transition.  It is believed that the real world has masses
which make the transition closer to
a crossover than a phase transition, but the evidence from lattice
Monte-Carlo studies is very weak.  In Fig. \ref{pisarski}, 
the various possibilities are shown.

\subsection{The Color Glass Condensate}

This section describes what is the Color Glass Condensate, 
and why it is important
for our understanding of basic properties of strong
interactions.\cite{cgc},\cite{cgc1} I
argue that the Color Glass Condensate is a universal form of matter which
controls the high energy limit of all strong interaction processes and is
the part of the hadron wavefunction important at such energies.  Since
the Color Glass Condensate is universal and controls the high energy limit
of all strong interactions,
it is of fundamental importance.
\subsubsection{What is the  Color Glass Condensate?}

The Color Glass Condensate is a new form of matter which controls
the high energy limit of strong interactions.  It is universal and
independent of the hadron which generated it. It should describe
\begin{itemize}
\item{ High energy cross sections}
\item{Distributions of produced particles}
\item{The distribution of the small x particles in a hadron}
\item{Initial conditions for heavy ion collisions}
\end{itemize}
Because this matter is universal, it is of fundamental interest.

A very high energy hadron has contributions to its wavefunction from gluons,
quarks and anti-quarks with energies up to that of the hadron 
and all the way down
to energies of the order of the scale of light mass hadron 
masses, $E \sim 200~MeV$.  A convenient variable in which to think
about these quark degrees of freedom is the typical energy of a
constituent scaled by that of the hadron,
\be
	x = E_{constituent}/E_{hadron}
\ee
Clearly the higher the energy of the hadron we consider, the lower is
the minimum $x$ of a constituent.  Sometimes it is also useful to consider
the rapidity of a constituent which is $y \sim ln(1/x)$

\begin{figure}[ht]
    \begin{center}
        \includegraphics[width=0.50\textwidth]{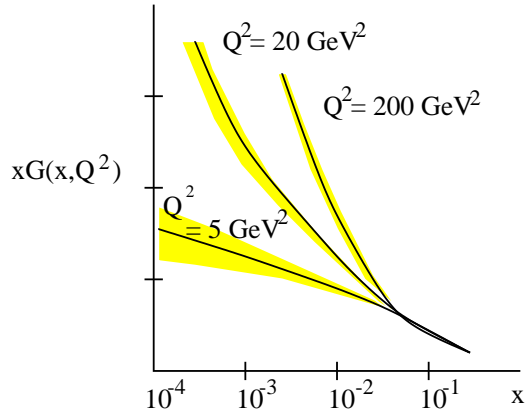}
        \caption{The number of gluons in a proton per unit rapidity
at various rapidities and $Q^2$ resolutions.}\label{gluon}
    \end{center}
\end{figure}

The density of small x partons is
\be
	{{dN} \over {dy}} = xG(x,Q^2)
\ee
The scale $Q^2$ appears because the number of constituents one measures
depends (weakly) upon the resolution scale of the probe with which one 
measures.  (Resolution scales are measured in units of the inverse momentum
of the probe, which is usually taken to be a virtual photon.)
A plot of $xG(x,Q^2)$ for gluons at various $x$ and $Q^2$
measured at the HERA accelerator in protons\cite{hera}, and
is shown in Fig. \ref{gluon}.

\begin{figure}[ht]
    \begin{center}
        \includegraphics[width=0.50\textwidth]{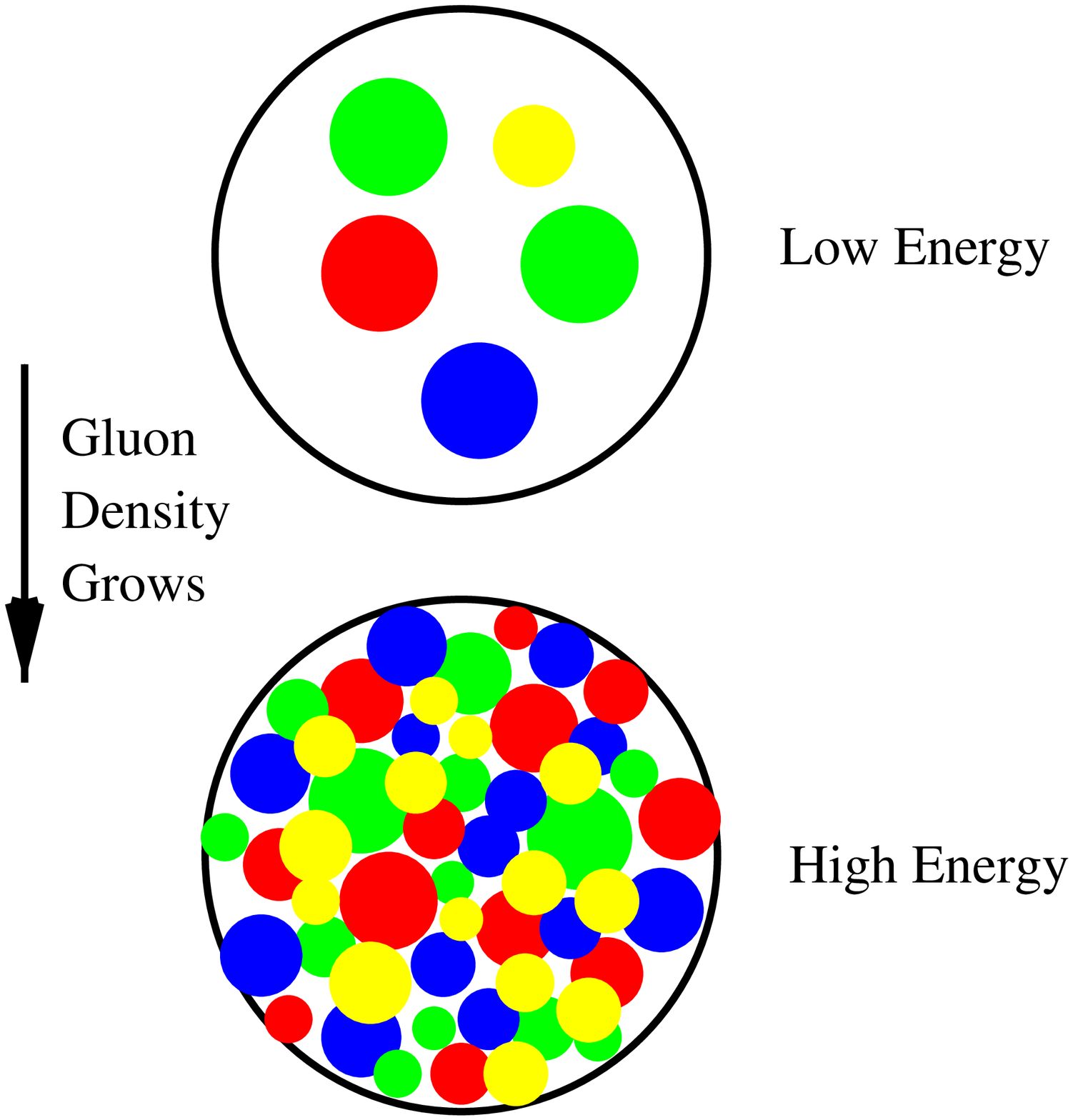}
        \caption{The increasing density of wee partons as
the energy increases.}\label{saturation}
    \end{center}
\end{figure}

Note that the gluon density rises rapidly at small x in Fig. \cite{hera}.
This is the so called small x problem.  It means that if we view the
proton head on at increasing energies, the low momentum gluon density 
grows.  This is shown in Fig.  \ref{saturation}.

As the density of gluons per unit area per unit rapidity increases, the 
typical transverse separation of the gluons decreases.  This means that the 
matter which controls high energy strong interactions is very dense, and it 
means that the QCD interaction strength, which is usually parameterized
by the dimensionless scale $\alpha_S$ becomes small.  The phase space density
of these gluons, $\rho \sim 1/\pi R^2 ~dN/d^2p_T$ can become at most $1/\alpha_s$
since once this density is reached gluon interactions are important.
This is  characteristic of Bose condensation phenomena which
are generated by an instability proportional to the density $\rho$
and is compensated by interactions proportional to $\alpha_S \rho^2$,
which become of the same order of magnitude when $\rho \sim 1/\alpha_s$
Thus the matter is a Color Condensate.

The glassy nature of the condensate arises because the fields associated
with the condensate are generated by constituents of the proton at higher
momentum.  These higher momentum constituents have their time scales 
Lorentz time dilated relative to those which would be measured in their rest
frame.  Therefore the fields associated with the low momentum constituents
also evolve on this long time scale.  The low momentum constituents are
therefore glassy: their time evolution scale is unnaturally long compared to 
their natural time scale.  Hence the name Color Glass Condensate.

There is also a typical scale associated with the Color Glass Condensate:
the saturation momentum.  This is the typical momentum scale where
the phase space density of gluons becomes $\rho \le 1/\alpha_S$.

\begin{figure}[ht]
    \begin{center}
        \includegraphics[width=0.35\textwidth]{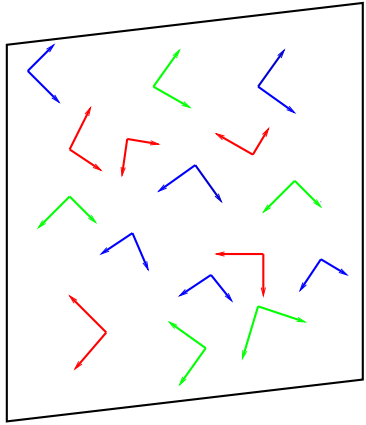}
        \caption{The Color Glass Condensate as a high density
of random gluon fields on a two dimensional sheet traveling near
the speed of light.}\label{colorglass}
    \end{center}
\end{figure}

At very high momentum, the fields associated with the Color Glass Condensate
can be treated as classical fields, 
like the fields of electricity and magnetism.
Since they arise from fast moving partons, they are plane polarized,
with mutually orthogonal color electric and magnetic fields
perpendicular to the direction of motion of the hadron.  They are also
random in two dimensions.  This is shown in Fig. \ref{colorglass}.

\subsubsection{Why is the Color Glass Condensate Important?}

\begin{figure}[ht]
    \begin{center}
        \includegraphics[width=0.60\textwidth]{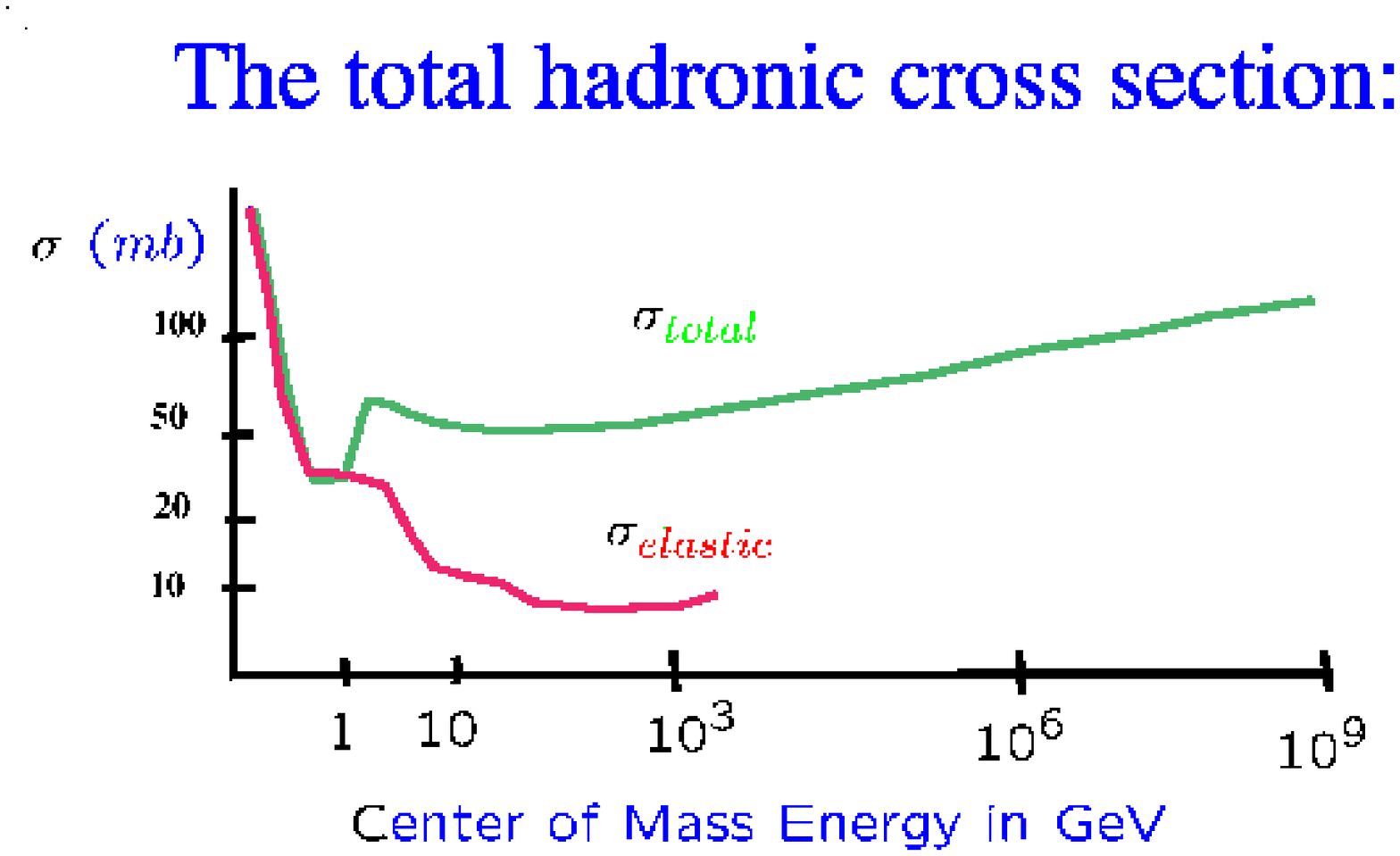}
        \caption{The total cross section for high energy
              proton-proton interactions.}\label{sigma}
    \end{center}
\end{figure}

Like nuclei
and electrons compose atoms, and nucleons and protons compose nuclear matter,
the Color Glass Condensate is the fundamental matter of which high energy
hadrons are composed. 
The Color Glass Condensate has the potential to allow for a 
first principles description of the gross or typical properties of
matter at high energies.  For example, the total cross section at high energies
for proton-proton scattering,
as shown in Fig. \ref{sigma} has a simple form but for over 40 years has 
resisted simple explanation.  
(It has perhaps been recently understood in terms of the Color Glass
Condensate or Saturation ideas.)\cite{Froissart}-\cite{kw}

\begin{figure}[ht]
    \begin{center}
        \includegraphics[width=0.40\textwidth]{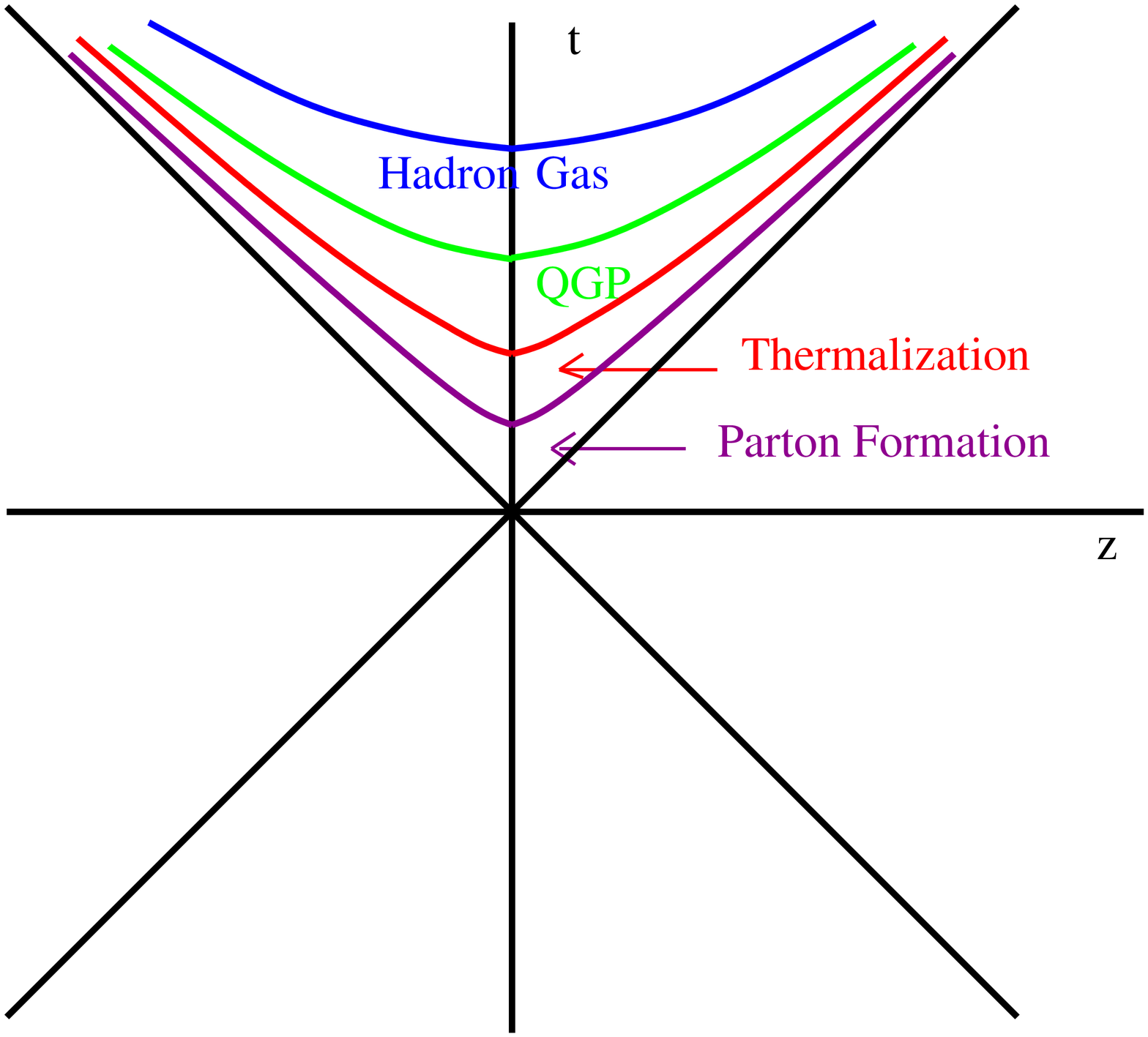}
        \caption{A space-time diagram for the evolution of
matter produced in heavy ion collisions.}\label{spacetime}
    \end{center}
\end{figure}

The Color Glass Condensate forms the matter in the quantum mechanical
state which describes a nucleus.  In the earliest stages of a nucleus-nucleus
collisions, the matter must not be changed much from these quantum mechanical 
states.  The Color Glass Condensate therefore provides the initial
conditions for the Quark Gluon plasma to form in these collisions.  A space-time
picture of nucleus nucleus collisions is shown in Fig. \ref{spacetime}.
At very early times, the Color Glass Condensate evolves into a distribution
of gluons.  Later these gluons thermalize and may eventually form a Quark
Gluon Plasma.  At even later times, a mixed phase of plasma and hadronic
gas may form.  

\section{Lecture II: Ultrarelativistic Nuclear Collisions.}

Heavy ion collisions at ultrarelativistic energies are visualized in Fig. 
\ref{sheetonsheet} as the collision of two sheets of colored glass.\cite{aa}

At ultrarelativistic energies, these sheets pass through one another.
In their wake is left melting colored glass, which eventually materializes as 
quarks and gluons.  These quarks and gluons would naturally form in their
rest frame on some natural microphysics time scale.  
For the saturated color glass, this
time scale is of order the inverse saturation momentum (again, we convert
momentum into time by appropriate uses of Planck's constant and the speed
of light), in the rest frame of the produced particle.  When a particle has
a large momentum along the beam axis, this time scale is Lorentz dilated.
This means that the slow particles are produced first towards the center of
the collision regions and the fast particles are produced later further away
from the collision region.  
\begin{figure}[ht]
    \begin{center}
        \includegraphics[width=.50\textwidth]{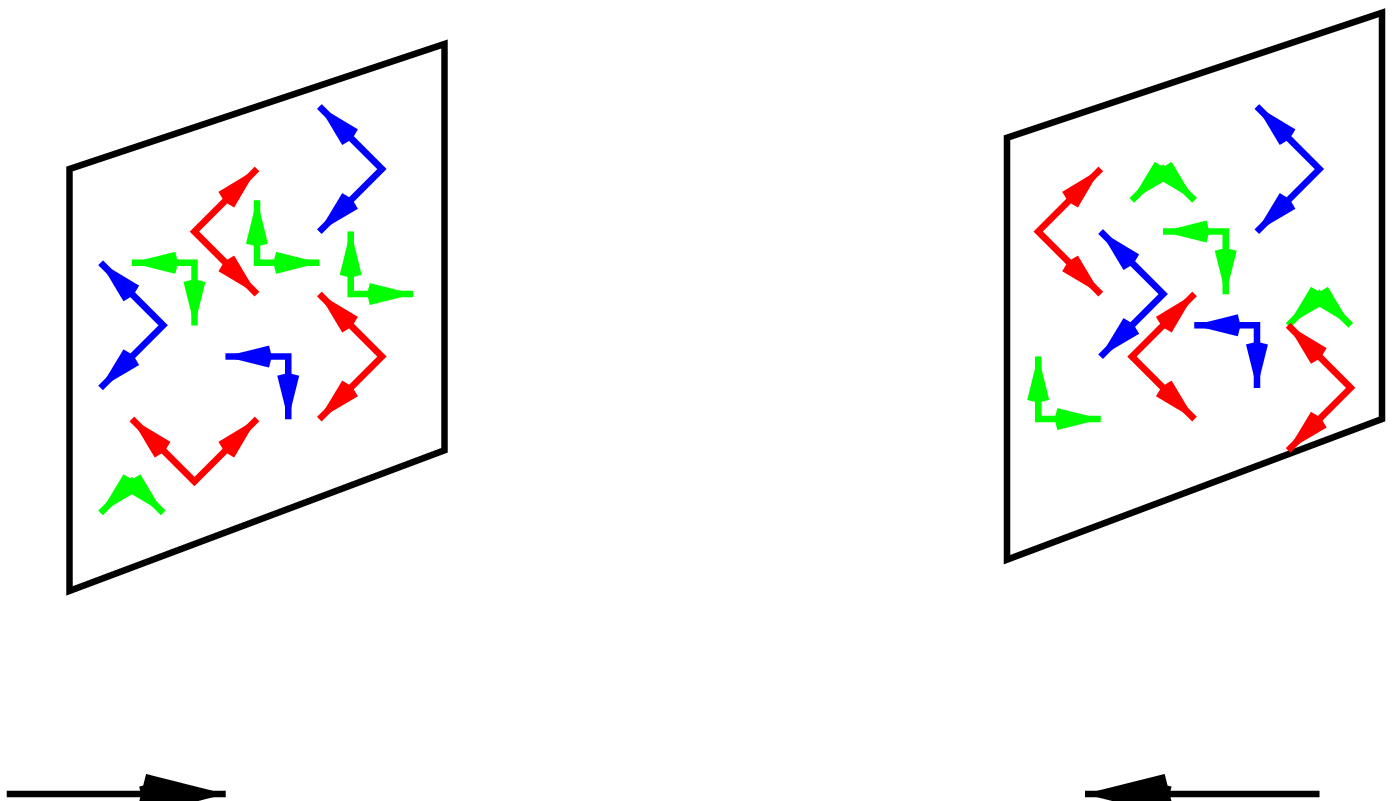}
        \caption{The collision of two sheets of colored 
glass.}\label{sheetonsheet}
    \end{center}
\end{figure}

\begin{figure}[ht]
    \begin{center}
        \includegraphics[width=.50\textwidth]{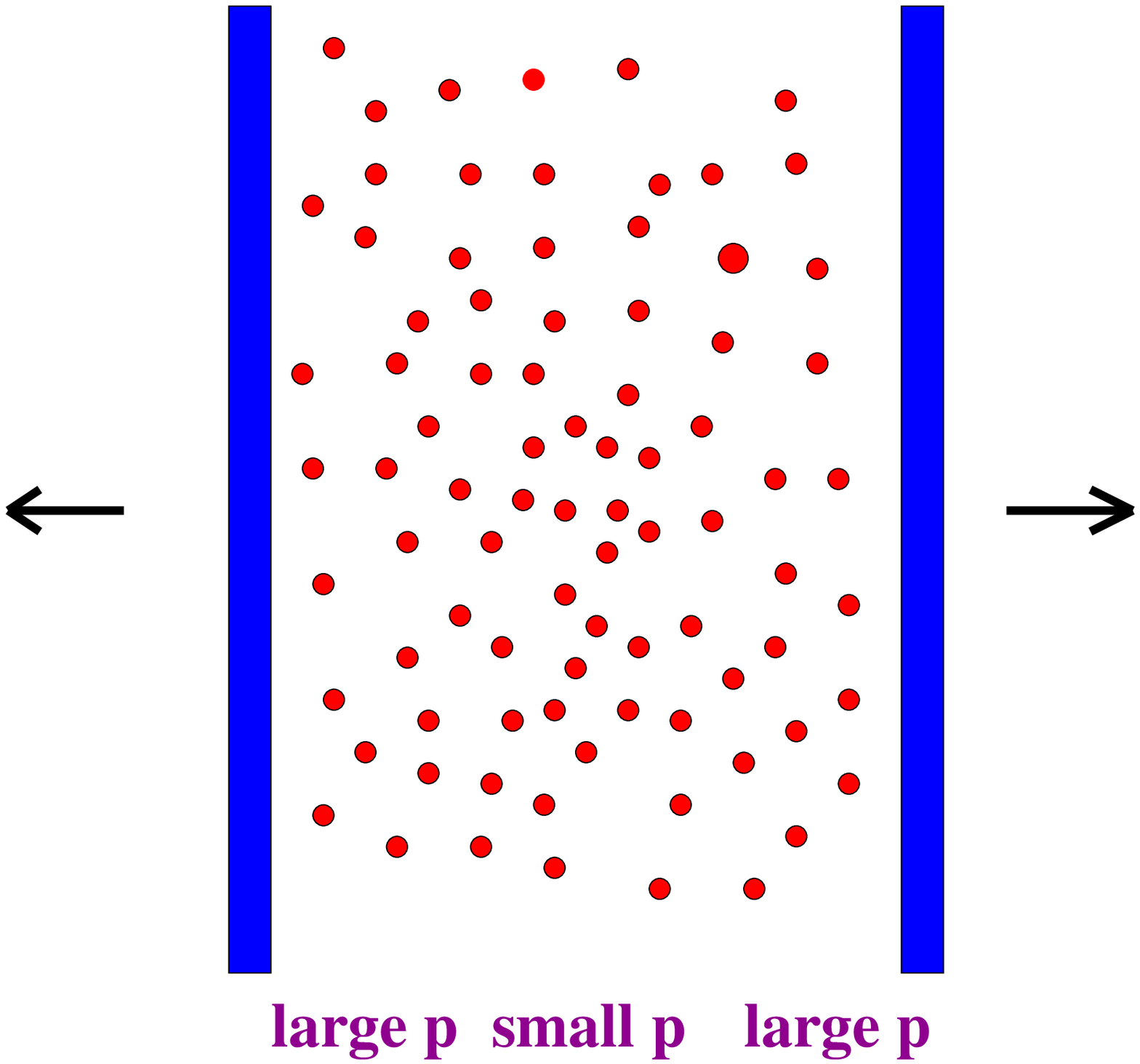}
        \caption{Particles being produced after the 
collision of two nuclei.}\label{collision}
    \end{center}
\end{figure}

This correlation between space and momentum is similar to what happens
to matter in Hubble expansion in cosmology.  The stars which are further away
have larger outward velocities.  This means that this system, like the universe
in cosmology is born expanding.  This is shown in Fig. \ref{collision}

As this system expands, it cools.  Presumably at some time the produced quarks
and gluons thermalize,  They then expand as a quark gluon plasma and eventually
as some mixture of hadrons and quarks and gluons.  Eventually, they may become
a gas of only hadrons before they stop interacting and fly off to detectors.

In the last lecture, we shall describe the results 
from nucleus-nucleus collisions
at RHIC in some detail. Before proceeding there, we need to 
learn a little bit more about the properties of high energy hadrons.
It is useful to introduce some kinematic variables which are useful
in what will follow.

The light cone momenta are defined as
\be
	P^{\pm} = {1 \over {\sqrt{2}}} (E \pm p_z)
\ee
and light cone coordinates are
\be
	X^{\pm} = {1 \over {\sqrt{2}}} (t \pm z)
\ee
The metric in these variables is
\be
	p \cdot x = p^+ x^- + p^- x^+ - p_T \cdot x_T
\ee
Conjugate variables are $x^\pm <-> p^\mp$.  The square of the four momentum
is
\be
	 p^2 = 2 P^+ P^- - P_T^2 = M^2
\ee
The uncertainty principle is
\be
	\Delta x^\pm \Delta p^\mp \ge 1
\ee

A reason why light cone variables are useful is because in a high
energy collision, a left moving particle has $p_Z \sim E$, so that
$p^+ \sim \sqrt{2} E$, but $p^- \sim m_T^2/p_z \sim 0$.  For the right
moving particles, it is $p^-$ which is big and $p^+$ which is very small.

Light cone variables scale by a constant under Lorentz transformations
along the collision axis.  Ratios of light cone momentum are therefore
invariant under such Lorentz boosts.  The light cone momentum fraction
$x = p^+_i/P^+$, where $P^+$ is that of the particle we probe
and $p_i^+$ is that of the constituent of the probed hadron satisfies
$0 \le x \le 1$.  It is the same as Bjorken x, and for a fast moving hadron,
it is almost Feynman $x_{Feynman} = E_i/E$.  This is the $x$ variable one is
using when one describes deep inelastic scattering.  In this case the label
$i$ corresponds to a quark or gluon constituent of a hadron.

One can also define a rapidity variable:
\be
	y = {1 \over 2} ln \left\{p_i^+ \over p_i^- \right\} \sim ln(2E_i/M_T)
\ee
Up to mass effects, the rapidity is in the range 
$-y_{proj} \le y \le y_{proj}$  When particles, like pions, are produced in
high energy hadronic collisions, one often plots them in terms of the
rapidity variable.  Distributions tend to be slowly varying functions
of rapidity.

\subsection{Is There Simple Behaviour at High Energy?}

\begin{figure}[ht]
    \begin{center}
        \includegraphics[width=.50\textwidth]{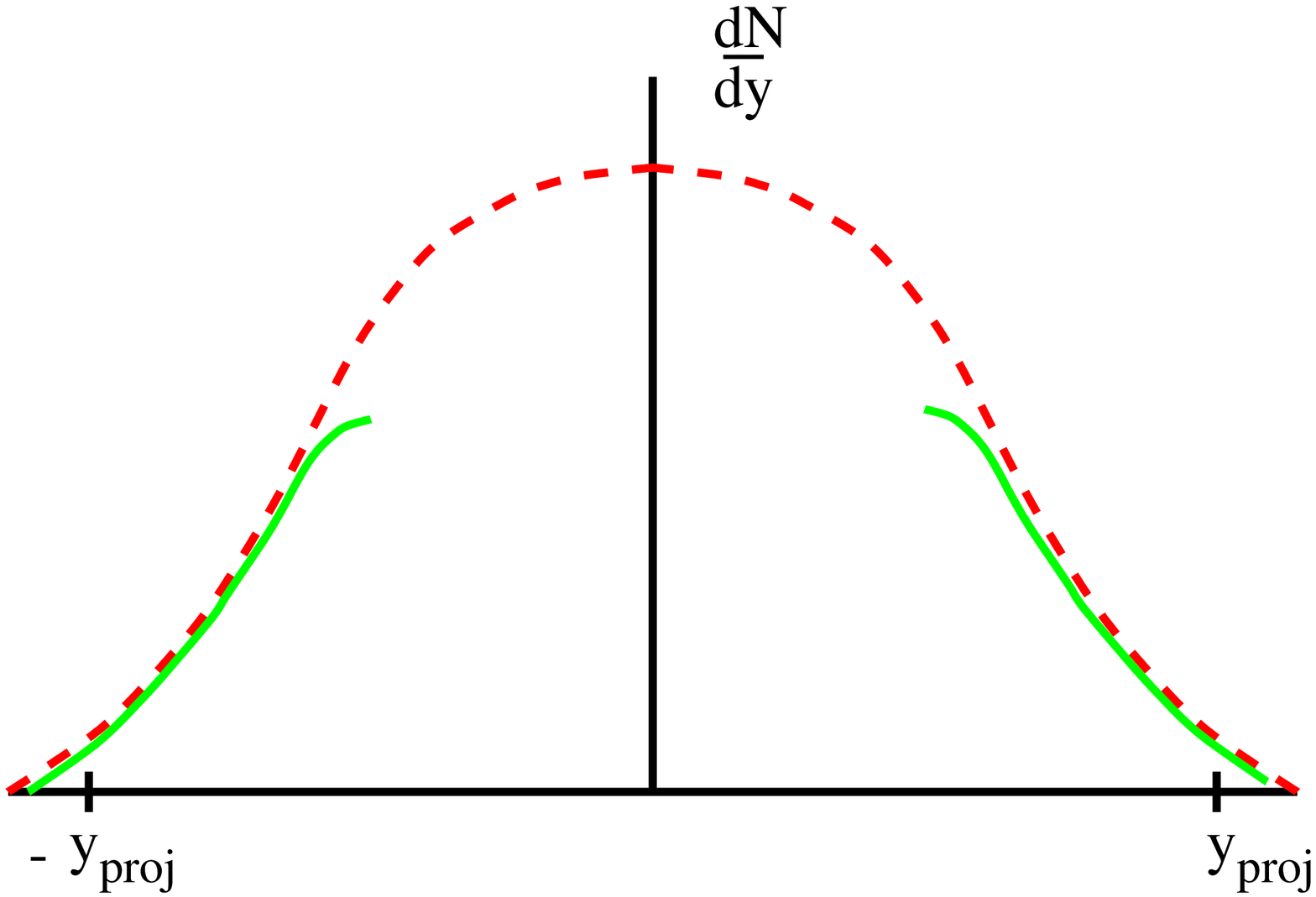}
        \caption{The rapidity distributions of particles at 
two different energies.}\label{feynman}
    \end{center}
\end{figure}

\begin{figure}[ht]
    \begin{center}
        \includegraphics[width=.50\textwidth]{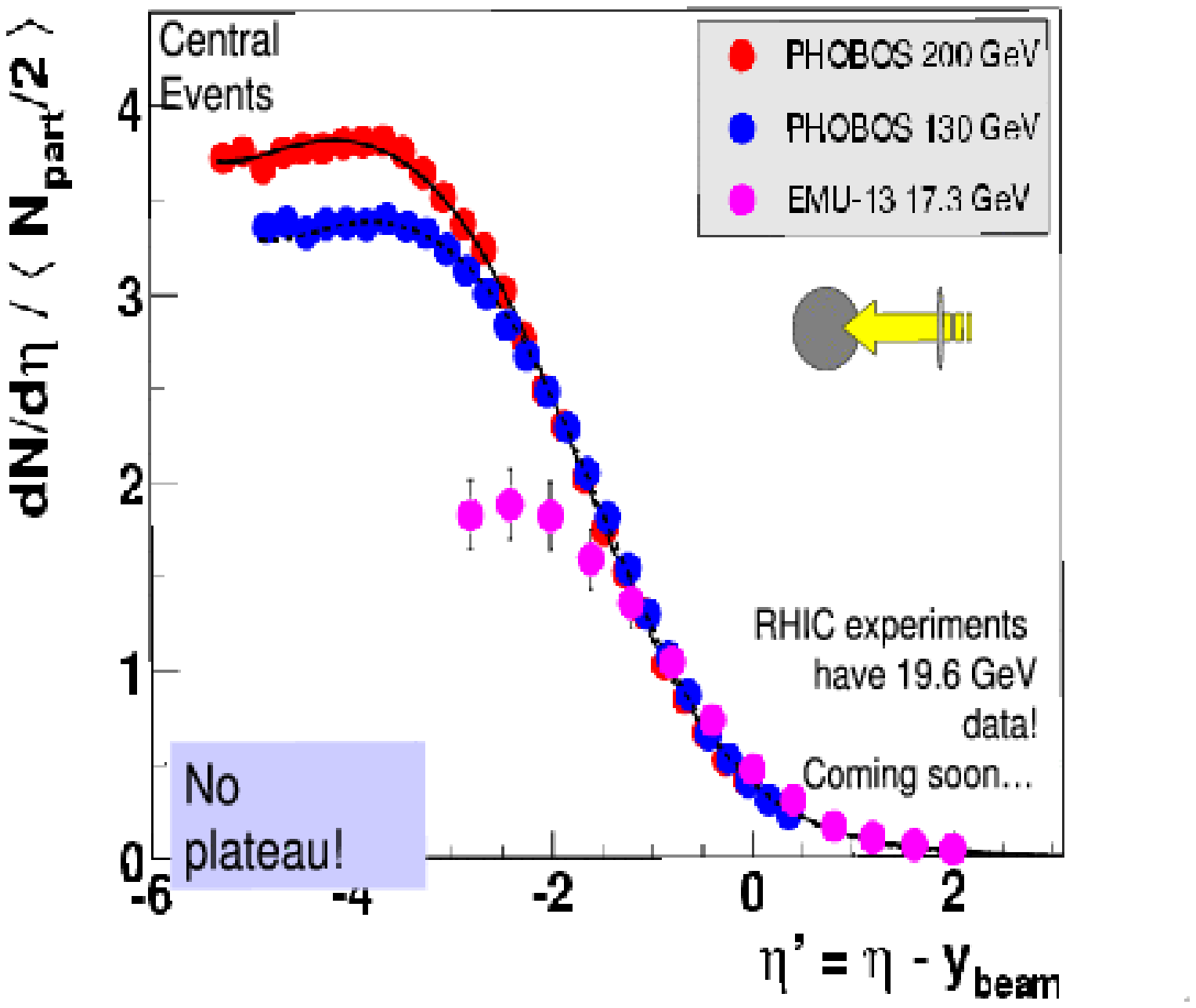}
        \caption{Experimental evidence from the Phobos experiment
at RHIC on limiting fragmentation.}\label{limitingfrag}
    \end{center}
\end{figure}

A hint of the underlying simplicity of high energy hadronic interactions
comes from studying the rapidity distributions of produced particles
for various collision energies.  In Fig. \ref{feynman},
a generic plot of of the rapidity distribution of produced pions is shown 
for two different energies.  The rapidity distribution at lower
energies has been cut in half and the particles associated with
each of the projectiles have been displaced in rapidity so that
their staring points in rapidity are the same.  It is remarkable, that except 
for the slowest particles in the center of mass frame, those 
located near $y \sim 0$, the distributions are almost
identical.\cite{limitingfrag}  
This is shown for the data from RHIC in Fig. \ref{limitingfrag}.

We conclude from this that going to higher energy adds in new degrees of
freedom, the small x part of the hadron wavefunction.  At lower energies,
these degrees of freedom are not kinematically relevant as they
can never be produced.  On the other hand, going to higher
energy leaves the fast degrees of freedom of the hadron unchanged.

This suggests that there should be a renormalization group description of
the hadrons.  As we go to higher energy, the high momentum degrees of
freedom remain fixed. Integrating out the previous small x degrees of
freedom should incorporate them into what are now the high energy 
degrees of freedom at the higher energy.  This process generates an
effective action for the new low momentum degrees of freedom.
Such a process, when done iteratively is a renormalization group.

\subsection{A Single Hadron}

\begin{figure}[ht]
    \begin{center}
        \includegraphics[width=.50\textwidth]{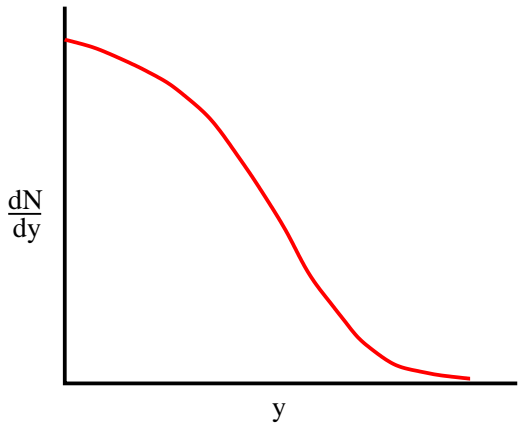}
        \caption{The rapidity distribution of the constituents of
a hadron.}\label{dndya}
    \end{center}
\end{figure}

A plot of the rapidity distribution of the
constituents of a hadron, the gluons, is shown generically in 
Fig. \ref{dndya}.
I have used $y = y_{hadron} -ln(1/x)$ as my definition
of rapidity.
This distribution is similar in shape to that of the half of the rapidity
distribution shown for hadron-hadron interactions in the center of mass frame
which has positive rapidity.  The essential difference is that
this distribution is for constituents where the hadron-hadron collision
is for produced particles, mainly pions.

\begin{figure}[ht]
    \begin{center}
        \includegraphics[width=.40\textwidth]{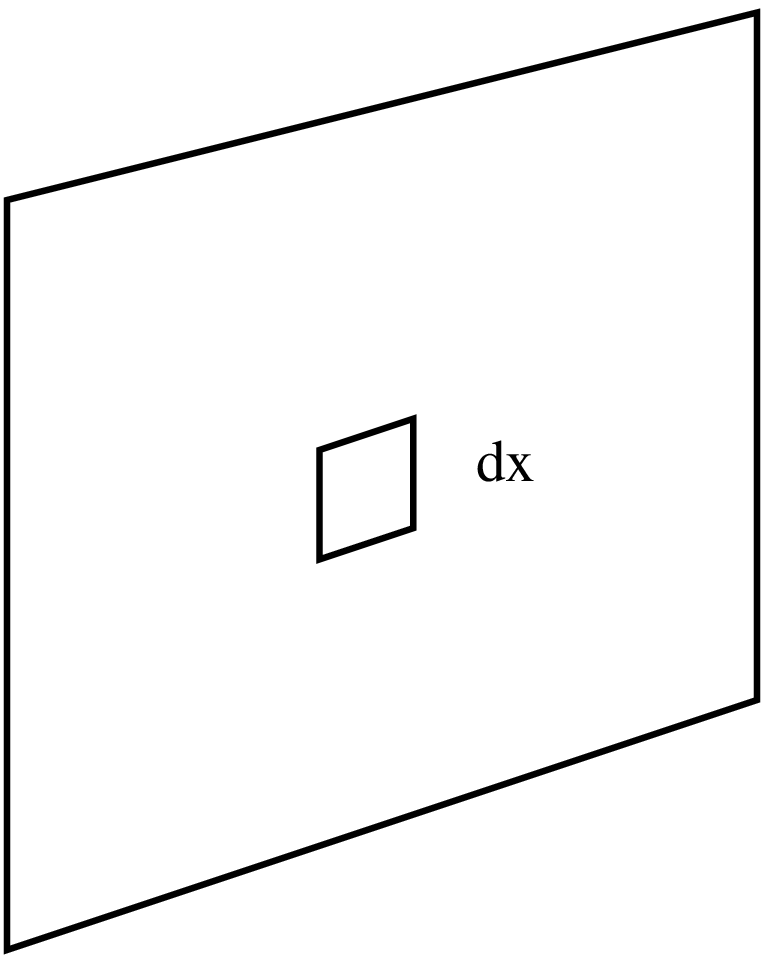}
        \caption{A thin sheet traveling near light velocity.
The transverse resolution scale is $\Delta x$}\label{sheet}
    \end{center}
\end{figure}

In the high energy limit, as discussed in the previous section,
the density of gluons grows rapidly.  This suggests we introduce a
density scale for the partons
\be
	\Lambda^2 = {1 \over {\pi R^2}} {{dN} \over {dy}}
\ee
One usually defines a saturation momentum to be $Q_{sat}^2 \sim \alpha_s 
\Lambda^2$, since this will turn out to be the typical momentum of particles
in this high density system. In fact, $\alpha_s$ is slowly varying compared
to the variation of $\lambda$, so that for the purposes of the estimates
we make here, whether or not there is a factor of $\alpha_s$ 
will not be so important.
Note that $\alpha_s$ evaluated at the saturation
scale will be $\alpha_s << 1$.  The typical particle transverse momenta are
of order $p_T^2 \sim Q_{sat}^2 >> 1/R_{had}^2$  Therefore it is consistent
to think of the parton distribution as a high density weakly coupled
system which is localized in the transverse plane. The high momentum partons,
the degrees of freedom which should be frozen, can be thought of as sitting on
an infinitesmally thin sheet.  We shall study this system with a resolution
size scale which is $\Delta x << 1/\Lambda_{QCD}$, so that we
may use weak coupling methods.  Such a thin sheet is shown in Fig. \ref{sheet}

It is useful to discuss different types of rapidity variables before 
proceeding.  The typical momentum space rapidity is
\be
	y & =& { 1\over 2} ln \left( p^+ \over p^- \right) \nonumber \\
         & = & ln \left( {2p^+} \over M_T \right) \nonumber \\
         & = & ln \left( {2p^+_{hadron}} \over M_T \right) + ln \left(
p^+ \over p^+_{hadron} \right) \nonumber \\
&= & y_{hadron} - ln(1/x)
\ee
Here $M_T$ is a particle transverse mass, and we have made approximations
which ignore overall shifts in rapidity by of order one unit.  Within these
approximations, the momentum space rapidity used to describe the production
of particles is the same as that used to describe the constituents of hadrons.

Oftentimes a coordinate space rapidity is introduced.  With 
$\tau = \sqrt{t^2 - z^2}$,
\be
	y = { 1 \over 2} ln \left( x^+ \over x^- \right) = ln(2\tau/x^-)
\ee
Taking $\tau$ to be a time scale of order $1/M_T$, and using
the uncertainty principle $x^\pm \sim 1/p^\mp$, we find that up
to shifts in rapidity of order one, all the rapidities are the same.
This implies that coordinate space and momentum space are highly correlated,
and that one can identify momentum space and coordinate space rapidity
with some uncertainty of order one unit.

\begin{figure}[ht]
    \begin{center}
        \includegraphics[width=.50\textwidth]{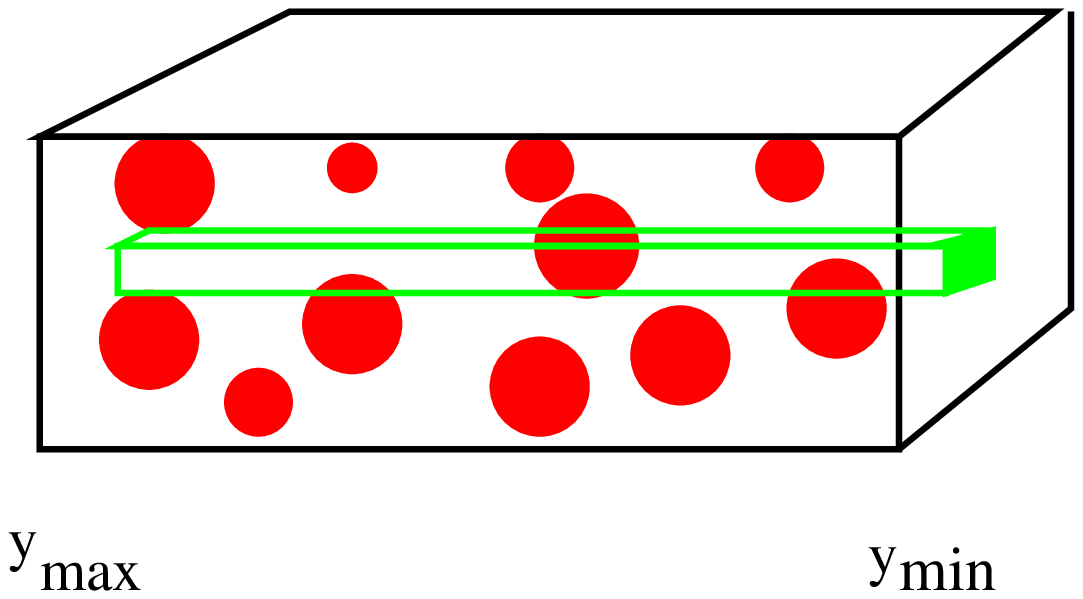}
        \caption{The distribution of particles
in a hadron in terms of rapidity variables.}\label{intersect}
    \end{center}
\end{figure}

If we plot the distribution of particles in hadron in terms of
the rapidity variable, the longitudinal dimension of the sheet
is spread out.  This is shown in Fig. \ref{intersect}.
The longitudinal position is correlated with the longitudinal momentum.
The highest rapidity particles are the fastest.  In ordinary
coordinate space, this means the fastest particles are those most
Lorentz contracted.  If we now look down a tube of 
transverse size $\Delta x << 1/\Lambda_{QCD}$, we will intersect
the various constituents of the hadron only occasionally.
The color charge probed by this tube should therefore be random,
until the transverse size scale becomes large enough so that it can probe
the correlations.  If the beam energy is large enough, or x is small enough,
there should be a large amount of color charge in each tube of fixed size 
$\Delta x$.  One can therefore treat the color charge classically.

The physical picture we have generated is that there should be
classical sources of to a good approximation random charges on a thin
sheet.  The current for this is
\be
	J^\mu_a = \delta^{\mu +} \delta (x^-) \rho_a (x_T)
\ee
The delta function approximation should be good for many purposes, but
it may also be useful in some circumstances to insert the
longitudinal structure
\be
	J^\mu_a = \delta^{\mu +} \rho_a(y,x_T)
\ee
and to remember that the support of the source is for very large $y$.

\subsection{The Color Glass Condensate}

We now know how to write down a theory to describe the Color Glass
Condensate.  It is given by the path integral\cite{cgc1}
\be
	\int [dA] [d\rho] \exp{\left(iS[A,\rho] - W[\rho] \right)}
\ee
Here $S[A,\rho]$ is the Yang-Mills action in the presence of a source
current as described above.  The function $W$ weights the various
configurations of color charge. In the simplest version of the Color
Glass Condensate, this can be taken to be a Gaussian
\be
	W = {1 \over 2}~\int dy d^2x_T~ {{\rho^2(y,x_T)} \over \mu^2(y)}
\ee
In this ansatz, $\mu^2(y)$ is the color charge squared density
per unit area per unit y scaled by $1/N_c^2-1$.  The theory
can be generalized to less trivial forms of the weight function, but
this form works at small transverse resolution scales,
$\Delta x << 1/Q_{sat}$.  As one increases
the transverse resolution scale one needs a better determination of $W$.
It turns out that at resolution scales of order $1/Q_{sat} <<
\Delta x << 1/\Lambda_{QCD}$, a Gaussian form  is still valid.

The averaging over an external field makes the theory of the Color Glass
Condensate similar to that of spin glasses.  The solutions of the
classical field equations also have $F^2 \sim 1/\alpha$, so the gluon fields
are strong and have high occupation number, hence the word condensate.

The theory described above has an implicit longitudinal momentum cutoff
scale.  Particles with momentum above this scale are treated as sources, and
those below it as fields.  One computes physical
quantities by first computing the classical fields
and then averaging over sources $\rho$.  This is a good approximation so
long as the longitudinal momentum in the field is not too far
below the longitudinal momentum cutoff, $\Lambda^+$.  If one
computes quantum corrections, the expansion parameter is
\be
	\alpha_s ln(\Lambda^+/p^+)
\ee

To generate a theory at smaller momenta, $\overline \Lambda^+$, one first
requires that $\alpha_s ln (\Lambda^+ / \overline \Lambda^+) <<1$.
Then one computes the quantum corrections in the presence of an the background
field.  This turns out to change only the weight function $W$.  Therefore
the theory maps into itself under a change of scale.  This is 
a renormalization group, and it determines the weight function 
$W$.\cite{cgc1},\cite{JKMW97}-\cite{JKLW97}

\subsection{Color Glass Fields}

The form of the classical fields is easily inferred.  On either side
of the sheet the fields are zero.  They have no time dependence, and in light
cone gauge $A^+ = 0$.  It is plausible to look for a solution which
is purely transverse.  On either side of the sheet, we have fields which are 
gauge transformations of zero field.  It can be a different
gauge transformation of zero field on different sides of the sheet.
Continuity requires that $F^{ij} = 0$.  $F^{i-}$ is zero because of
light cone time $x^+$ independence, and the assumption that
$A^- = 0$.  $F^{i+}$ is non zero $\sim \delta (x^-)$ because of the
variation in $x^-$ as one crosses the sheet.  This means that
$F^{i0} \sim - F^{iz}$, or that $E \perp B \perp \vec{z}$.
These are transversely polarized Weiszacker-Williams fields.  They are random
in the two dimensional plane because the source is random.  This is shown in 
Fig.  \ref{colorglass}.
The intensity of these fields is of order $1/\alpha_s$, and they are 
not at all stringlike.

\subsection{The Gluon Distribution and Saturation}

The gluon distribution function is given by computing
the expectation value of the number operator $<a^\dagger(p) a(p)>$
and can be found from computing the gluon field
expectation value $<A(p) A(-p) >$.  This is left as an exercise
for the student.  At large $p_T$, the distribution function scales as
\be
	{{dN} \over {dyd^2p_T}} \sim {1 \over \alpha_s}~
{Q_{sat}^2 \over p_T^2}
\ee which is typical of a Bremstrahlung spectrum  At small $p_T$,
the solution is $\sim ln(Q_{sat}^2/p_T^2) /\alpha_s$.  The reason for
this softer behaviour at smaller $p_T$ is easy to understand.  At
small distances corresponding to large $p_T$, one sees point sources of charge,
but at smaller $p_T$, the charges cancel one another and lead to a
dipole field.  The dipole field is less singular at large $r$, and when
transformed into momentum space, one loses two powers of momentum
in the distribution function.  In terms of 
the color field, the saturation
phenomena is almost trivial to understand.  (It is very difficult
to understand if the gluons are treated as incoherently
interacting particles.)

Now $Q_{sat}^2$ can grow with energy. In fact it turns out that
$Q_{sat}^2$ never stops growing.  The intrinsic transverse
momentum grows without bound.  Physically what is happening is that
the low momentum degrees of freedom below the saturation momentum
grow very slowly, like $ln(Q_{sat}^2)$ because repulsive gluon
interactions prevent more filling.  On the other hand,
one can always add more gluons at high momentum since the phase
space is not filled there.

How is this consistent with unitarity?  Unitarity is a statement
about cross sections at fixed $Q^2$.  If $Q^2$ is above the saturation
momentum, then the gluon distribution function grows rapidly with energy,
as $Q_{sat}^2$.  On the other hand, once the saturation momentum becomes
larger than $Q^2$, the number of gluons one can probe
\be
	xG(x,Q^2) \sim \pi R^2 \int_0^{Q^2} d^2p_T ~ {{dN} \over {d^2p_Tdy}}
\ee 
varies only logarithmically.  The number of gluons scale as
the surface area.  (At high $Q^2$, it is proportional to $R^2 Q_{sat}^2$,
and one expects that $Q_{sat}^2 \sim A^{1/3}$ so that $xG(x,Q^2) \sim A$

\subsection{Hadron Collisions}
 
In Fig. \ref{sheetonsheet}, the collision of two hadrons is
\begin{figure}[ht]
    \begin{center}
        \includegraphics[width=.50\textwidth]{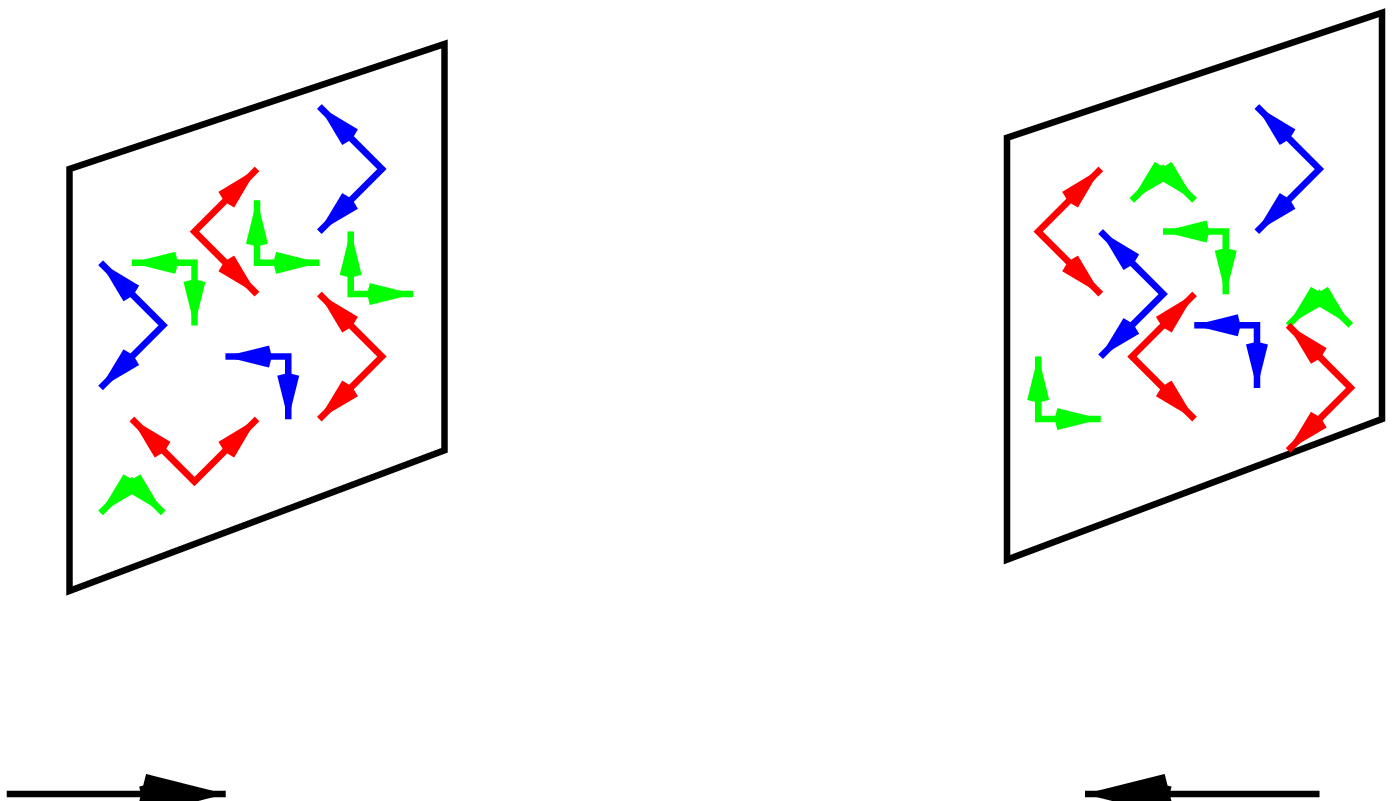}
        \caption{The collision of two sheets of colored glass.}
\label{sheetonsheet}
    \end{center}
\end{figure}
represented as that of two sheets of colored glass.  Before the
collisions, the left moving hadron has fields
\be
  F^{i+} & \sim & \delta(x^-) \nonumber \\
  F^{ij} & \sim & 0 \nonumber \\
  F^{i-} & \sim & 0 \\
\ee
and that of the right moving fields is analogous to 
that of the above save that $\pm \rightarrow \mp$ in the indices
and coordinates of all fields.  The fields are of course different
in each nucleus.  We shall consider impact parameter zero head on
collisions in what follows.

These fields are plane polarized and have random colors.
A solution of the classical Yang-Mill's equation can be constructed
by requiring that the fields are two dimensional gauge transforms
of vacuum everywhere but in the forward light cone.  At the edges of the
light cone, and at its tip $t = z =0$, the equations are singular,
and a global solution requires that the fields carry non-trivial
energy and momentum in the forward lightcone.  At short times,
these
fields are highly non-linear. In a time of order $\tau \sim 1/Q_{sat}$,
the fields linearize.  When they linearize, we can identify
the particle content of the classical radiation field.  

This situation is much different than the case for quantum electrodynamics.
Because of the gluon self-interaction, it is possible to classically
convert the energy in the incident nuclei into radiation.  In quantum
electrodynamics, the charged particles are fermions, and they cannot
be treated classically.  Radiation is produced by annihilation
or bremstrahlung as quantum corrections to the initial value problem.

The solution to the field equation in the forward lightcone is 
approximately boost invariant over an interval of rapidity of
order $\Delta y \ll 1/\alpha_s$.  At large momentum, the field
equations can be solved in perturbation theory and the distribution is
like that of a bremstrahlung spectrum 
\be
   {{dN} \over  {dyd^2p_T}} \sim {1 \over \alpha_s} \pi R^2 
{Q_{sat}^4 \over p_T^4}
\ee
It can be shown that such a spectrum matches smoothly onto the result
for high momentum transfer jet production.

One of the outstanding problems of particle production is computing the total
multiplicity of produced gluons.  In the CGC description, this problem is 
solved.  When $p_T \le Q_{sat}$, non-linearities of the field
equations become important, and the field stops going as $1/p_T^4$.
Instead it becomes of order 
\be
	{{dN} \over {dyd^2p_T}} \sim { 1 \over \alpha_s} \pi R^2 
\ee
The total multiplicty is therefore of order
\be
	N \sim {1 \over \alpha_s} ~\pi R^2 Q_{sat}^2
\ee

If $Q_{sat}^2 \sim A^{1/3}$, then the total multiplicty goes as A,
the high $p_T$ differential multiplicity goes as $A^{4/3}$, as we naively 
expect for hard processes since they
should be incoherent , and the low momentum differential multiplicity
goes as $A^{2/3}$, since these particles arise from the region where
the hadrons are black disks and the emission should take place from the 
surface.  

\begin{figure}[ht]
    \begin{center}
        \includegraphics[width=.50\textwidth]{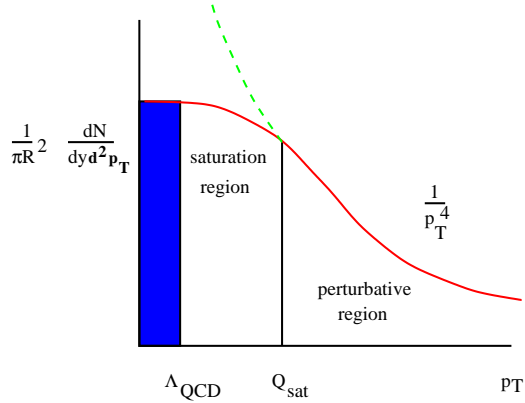}
        \caption{A cartoon representation of the various
kinematic regions of gluon production.}\label{pt4}
    \end{center}
\end{figure}

\begin{figure}[ht]
    \begin{center}
        \includegraphics[width=.50\textwidth]{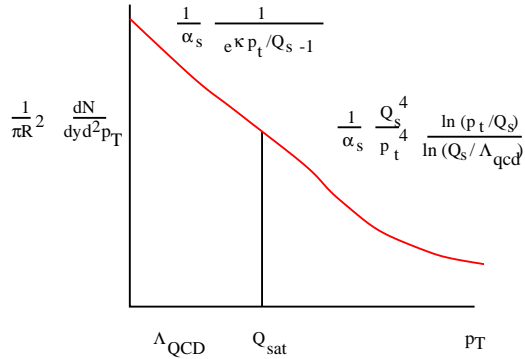}
        \caption{The numerically computed distribution
of produced gluons.}\label{ptraju}
    \end{center}
\end{figure}

In Fig. \ref{pt4}, the various kinematic regions for production of gluons are
shown.
In Fig. \ref{ptraju}, the results of numerical simulation
of gluon production are shown.
At small $p_T$, it is amusing that the distribution is well
described by a two dimensional Bose-Einstein distribution.  This is
presumably a numerical accident, and in this computation has absolutely
nothing to do with thermalized distributions.

\subsection{Thermalization}

As shown in Fig. \ref{collision}, in a heavy ion collision, the 
slow particles are produced first near the collision point and
the slow particles are produced later far from the collision point.
This introduces a gradiant into the initial matter distribution,
and the typical comoving volume element expands like $1/\tau$.
To understand the factor of $1/\tau$ in the above equation, note that
if we convert $dN/dz = dN/dy ~dy/dz = dN/dy ~1/t$, where we used our previous 
definition of space-time rapidity, and where we evaluated at
$z=0$.  This is the physical rest frame density at $z=0$.

If entropy is conserved, as is the case for thermalized system
with expansion time small compared to collision time,
\be
	S \sim T^3 \tau R^2
\ee
is fixed so that $T \sim 1/\tau^{1/3}$.  Therefore for a thermalized
system, the energy density decreases as $\epsilon \sim 1/\tau^{4/3}$,
where for system with no scattering so that the typical transverse
momentum does not change, $\epsilon \sim 1/\tau$.

For the initial
conditions typical of a Color Glass Condensate, 
thermalization is not so easy to do.\cite{BMSS} 
At the earliest times, the typical transverse momentum
is large, of order of the saturation momentum.  At this
scale, the coupling is weak $\alpha_s(Q_{sat}) << 1$, at least
for asymptotically large energy.  

To estimate the typical scattering time, we need to know the density
and the mean free path.  At early times, the density is that in the transverse
space diluted by the longitudinal expansion of the system,
\be
	\rho = <p_T^2> /\tau
\ee
The scattering cross section is on the other hand $\sigma \sim
\alpha^2_s ln(\rho) /<p_T^2>$. The logarithmic cutoff
comes about from Debye screening the Coulomb cross section.
(The linear divergence can be shown to cancel for thermalization
processes.)

Thermalization requires that $\tau >> \tau_{scat}$, since $\tau$
itself is the characteristic expansion time.  This requires that
\be
         \tau \ge exp(c/\alpha_s) 1/Q_{sat}
\ee
For practical purposes and for weakly coupled systems, 
there is never thermalization by elastic 
scattering!   

Thermalization, if it in fact occurs, takes place by inelastic scattering.
The physics of what is happening is easy to understand.  Because the system
begins its evolution with $p_T$ at such a large typical scale $Q_{sat}$,
the coupling is weak and the system does not easily thermalize by elastic 
scattering.  It therefore expands and becomes an overly dilute compared
to the typical density associated with the transverse 
momentum scale $p_T^3$.  When a system is overly
dilute, the Debye screening length becomes very large.  Multigluon
production processes can be shown to diverge like
the Debye screening length, whereas elastic processes only
diverge like the logarithm of this length.  Therefore, when
the Debye screening length is of order $1/\alpha_s$, multigluon
production begins to become more important than elastic
scattering.  This happens at a time $\tau \sim 1/(\alpha_s Q_{sat})$.

The details of how this thermalization occurs have not been fully
worked out in detail.  Current estimates of the time of thermalization 
matter produced in heavy ion collisions 
at RHIC energies ranges from $.3 \le \tau \le 3 ~Fm/c$.

\section{Lecture 3: What We Have Learned from RHIC}

In this lecture, I review results from RHIC and describe what we have so far
learned about the production of new forms of matter in heavy ion collisions.
I will make the case that we have produced matter of extremely high energy 
density, so high that it is silly not to think of it
as composed of quarks and gluons.  I also will argue that this matter is
strongly interacting with itself.  The issue of the properties of this matter
is still largely unresolved.  For example whether the various 
quantities measured
are more properly described as arising from a Color Glass Condensate or
from a Quark Gluon Plasma, although we can easily understand in most cases 
which form of matter should be most important.

\subsection{The Energy Density is Big}

\begin{figure}[h]
    \begin{center}
        \includegraphics[width=0.80\textwidth]{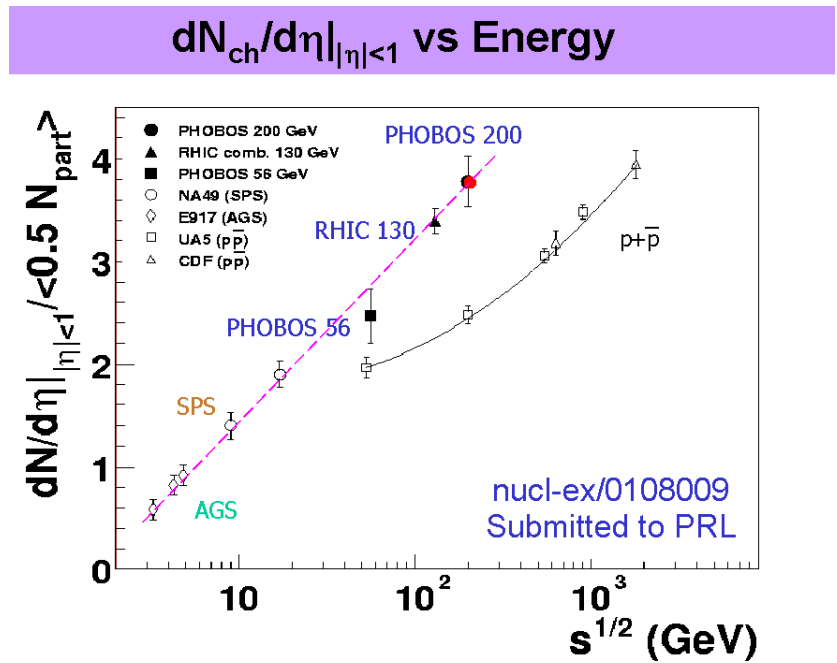}
        \caption{The particle multiplicity as a function of energy
as has been measured at RHIC.  
                            }\label{dndye}   
 \end{center}
\end{figure}

The  particle multiplicity as a function of energy has been measured at 
RHIC, as shown in Fig. \ref{dndye}. 
Combining the multiplicity data together with the measurements of transverse
energy or of typical particle transverse momenta, one can determine the
energy density of the matter when it decouples.\cite{dndyphobos}  
One can then extrapolate
backwards in time.  We extrapolate backwards using 1 dimensional 
expansion, since decoupling occurs when the matter first begins
to expands three dimensionally.  We can extrapolate backwards until  
the matter has melted from a Color Glass.

To do this extrapolation we use that the density
of particles falls as $N/V \sim 1/t$ during 1 dimensional expansion.  
If the particles expand without
interaction, then the energy per particle is constant.  If the particles
thermalize, then $E/N \sim T$, and since $N/V \sim T^3$ for a massless
gas, the temperature falls as $T \sim t^{-1/3}$.  For a gas which is not
quite massless, the temperature falls somewhere in the range $T_o
> T > T_o (t_o/t)^{1/3}$, that is the temperature is bracketed
by the value corresponding to no interaction and to that of a massless 
relativistic gas.  This 1 dimensional expansion continues
until the system begins to feel the effects of finite size in the 
transverse direction, and then rapidly cools through three dimensional
expansion.  Very close to when three dimensional expansion begins, the
system decouples and particles free stream without 
further interaction to detectors.
\begin{figure}[h]
    \begin{center}
        \includegraphics[width=0.80\textwidth]{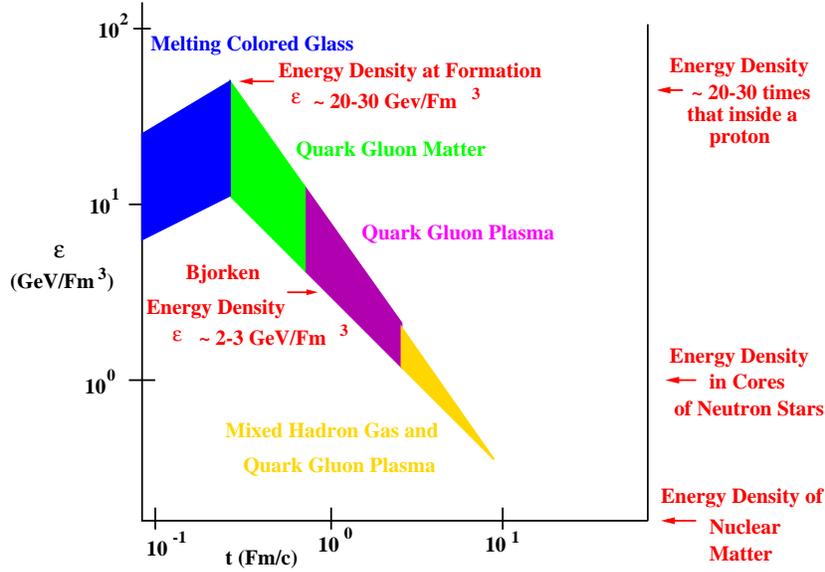}
        \caption{Bounds on the energy density as a function of time
in heavy ion collisions.  
                            }\label{times}   
 \end{center}
\end{figure}
We shall take a conservative 
overestimate of this time to be of order $t_{melt} \sim .3 ~Fm/c$  
The
extrapolation backwards is bounded by $\epsilon_{f} (t_f/t)
< \epsilon(t) < \epsilon_f (t_f/t)^{4/3}$.  The lower bound is
that assuming that the particles do not thermalize and their typical
energy is frozen.  The upper bound assumes that the system thermalizes
as an ideal massless gas.  We argued above that the true result is 
somewhere in between.  When the time is as small as the melting time,
then the energy density begins to decrease as time is further decreased.

This bound on the energy density is shown in Fig. \ref{times}.
On the left axis is the energy density and on the bottom axis is time.
The system begins as a Color Glass Condensate, then melts to Quark
Gluon Matter which eventually thermalizes to a Quark Gluon Plasma.
At a time of a few $Fm/c$, the plasma becomes a mixture of quarks, gluons and 
hadrons which expand together.  

At a time of about $10 ~Fm/c$, the
system falls apart and decouples.  At a time of $t \sim 1~Fm/c$,
the estimate we make is identical to the Bjorken energy density estimate,
and this provides a lower bound on the energy density achieved 
in the collision.  (All estimates agree that by a time of order $1 ~Fm/c$,
matter has been formed.)  The upper bound corresponds to assuming
that the system expands as a massless thermal gas from a melting
time of $.3~Fm/c$.  (If the time was reduced, the upper bound would  
be increased yet further.)
The bounds on the energy density 
are therefore
\be
	2-3 ~GeV/Fm^3 < \epsilon < 20-30 ~GeV/Fm^3
\ee
where we included a greater range of uncertainty in the upper limit
because of the uncertainty associated with the formation time.
The energy density of nuclear matter is about $0.15~GeV/Fm^3$, and
even the lowest energy densities in these collisions is in excess of this.
At late times, the energy density is about that of the cores of neutron stars,
$\epsilon \sim 1 ~GeV/Fm^3$.

{\bf At such extremely high energy densities, 
it is silly to try to describe the
matter in terms of anything but its quark and gluon degrees of freedom.}

\subsection{The Gross Features of Multiplicity Distributions Are Consistent
with Colored Glass}

\begin{figure}[ht]
    \begin{center}
        \includegraphics[width=.70\textwidth]{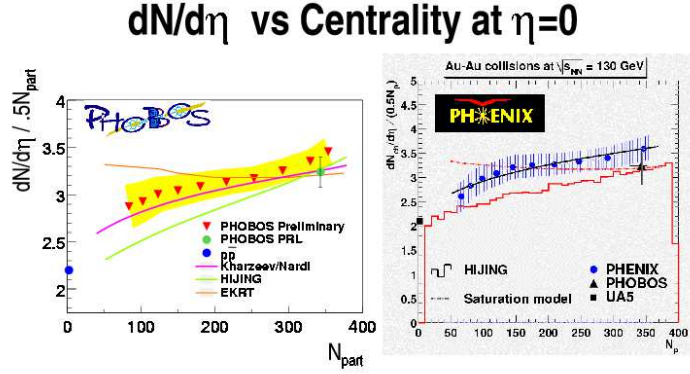}
        \caption{The CGC description of the participant
dependence of the multiplicity of produced particles.}\label{npart}
    \end{center}
\end{figure}

\begin{figure}[htb]
    \centering
       \mbox{{\epsfig{figure=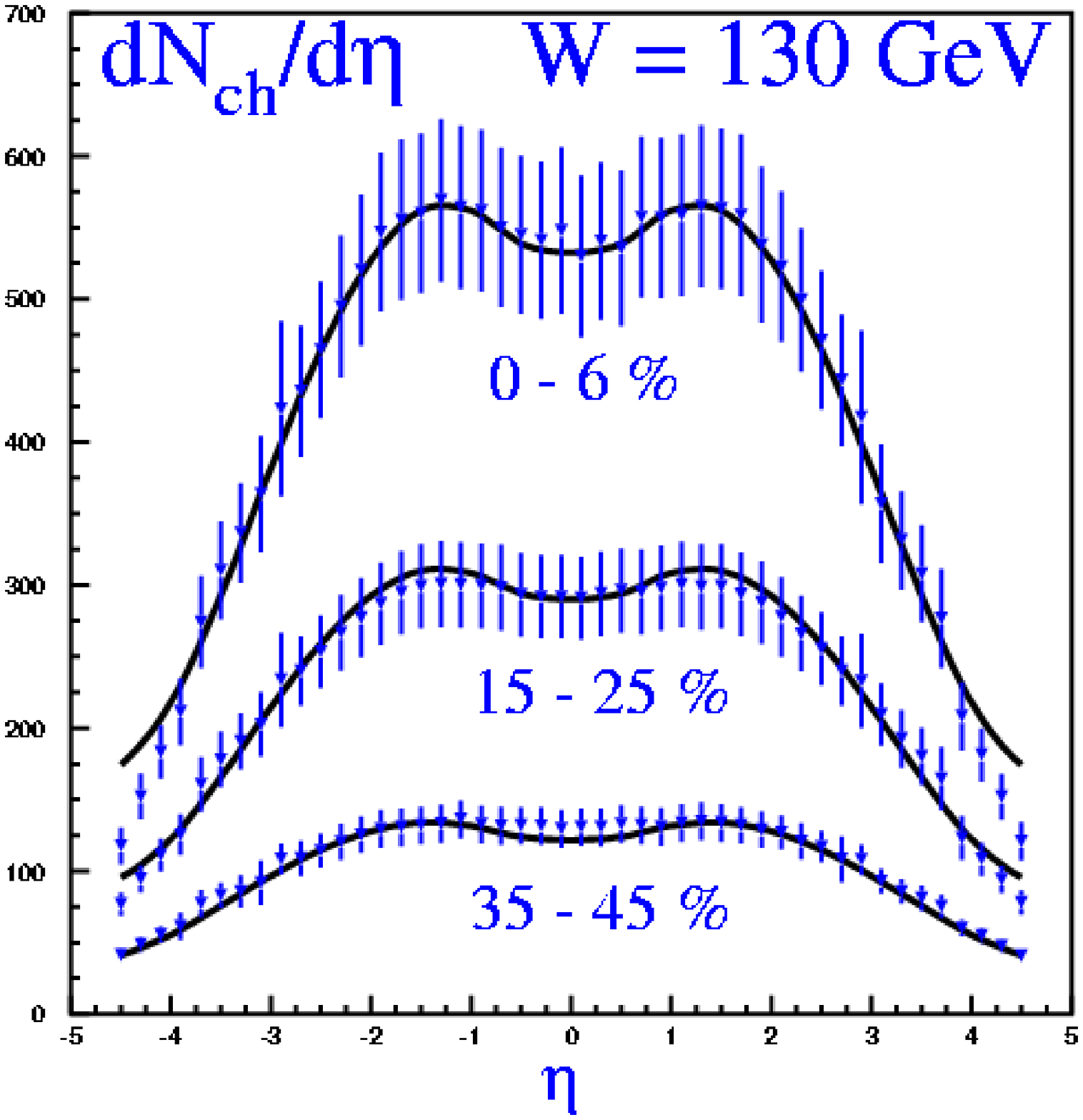,
        width=0.50\textwidth}}\quad
             {\epsfig{figure=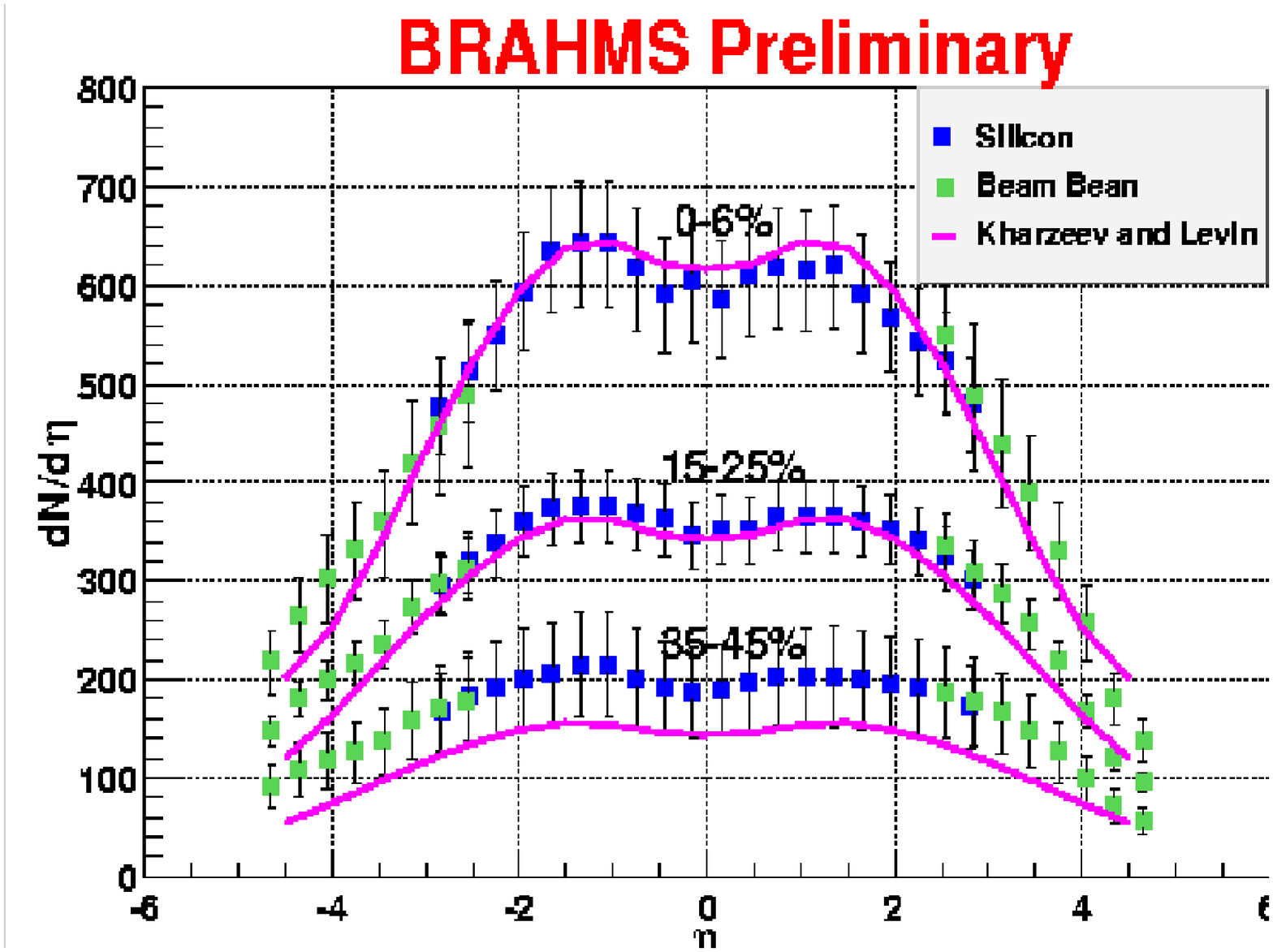,   
        width=0.50\textwidth}}}
        \caption{Color glass condensate fits to the rapidity density
measured in the PHOBOS and Brahms experiments}
        \label{dndysat}
\end{figure}

As argued by Kharzeev and Nardi,\cite{KN} the density of produced
particles per nucleon which participates in the collision,
$N_{part}$,  should be proportional to
$1/\alpha_s(Q_{sat})$, and $Q_{sat}^2 \sim N_{part}$. This dependence 
follows from the $1/\alpha_s$ which characterizes the density of
the Color Glass Condensate.
In Fig. \ref{npart}, we show the 
data from PHENIX and PHOBOS\cite{npart}.  The 
Kharzeev-Nardi form fits the data well.  Other attempts such as 
HIJING\cite{hijing},
or the so called saturation model of 
Eskola-Kajantie-Ruuskanen-Tuominen\cite{ekrt} are also shown in the
figure.

Kharzeev and Levin have recently argued that the rapidity distributions
as a function of centrality can be computed from the 
Color Glass description.\cite{kl}  This is shown in 
Fig. \ref{dndysat}.\cite{phbr}

\subsection{Matter Has Been Produced which Interacts Strongly 
with Itself}

\begin{figure}[htb]
    \begin{center}
        \includegraphics[width=0.80\textwidth]{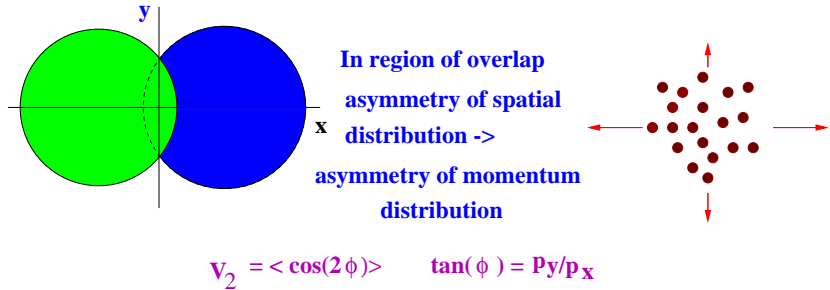}
        \caption{The asymmetry in the distribution of matter in an off center
collision is converted to an asymmetry of the momentum space distribution.  
                            }\label{flow}   
 \end{center}
\end{figure}

\begin{figure}[htb]
    \centering
       \mbox{{\epsfig{figure=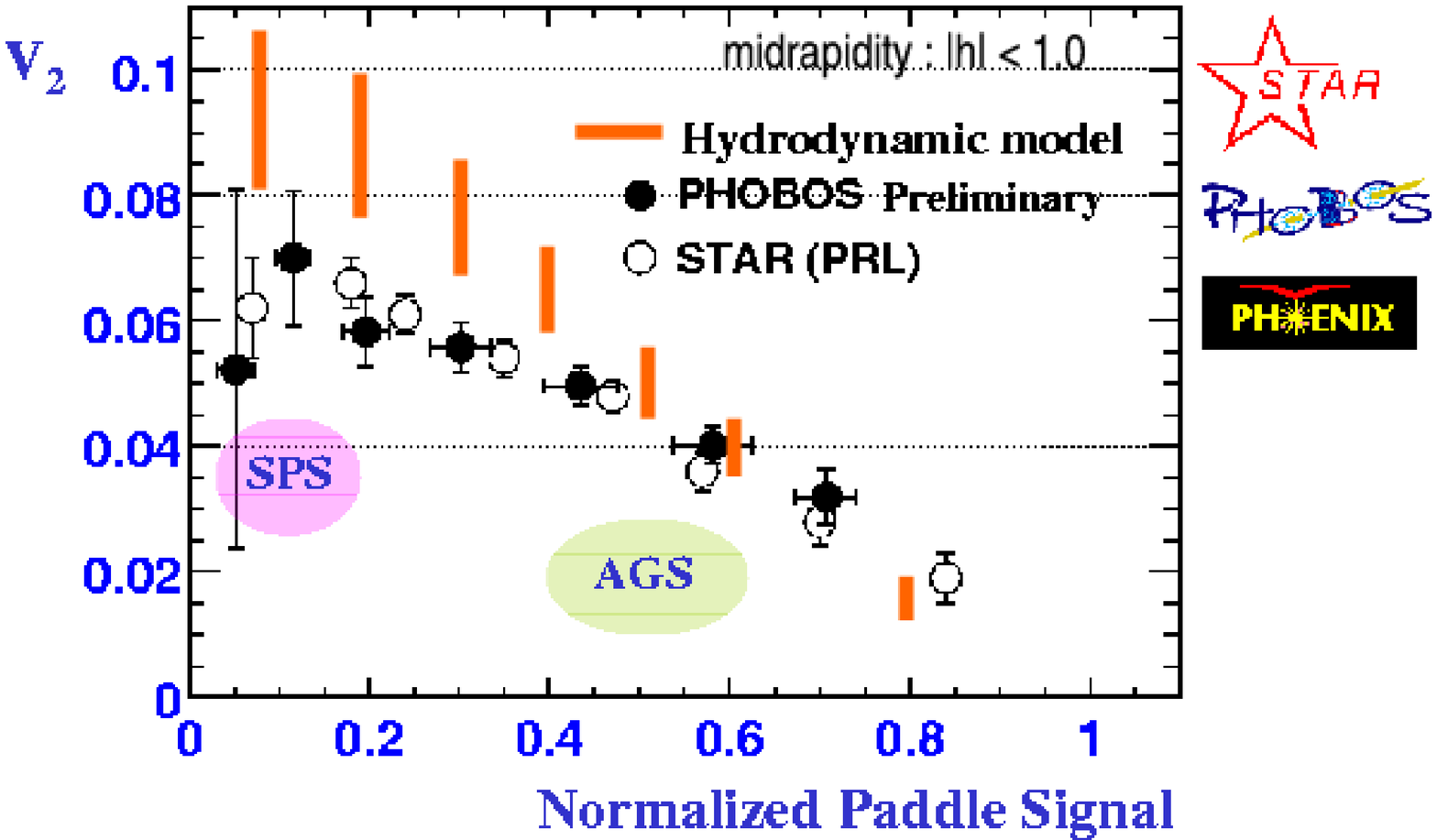,
        width=0.50\textwidth}}\quad
             {\epsfig{figure=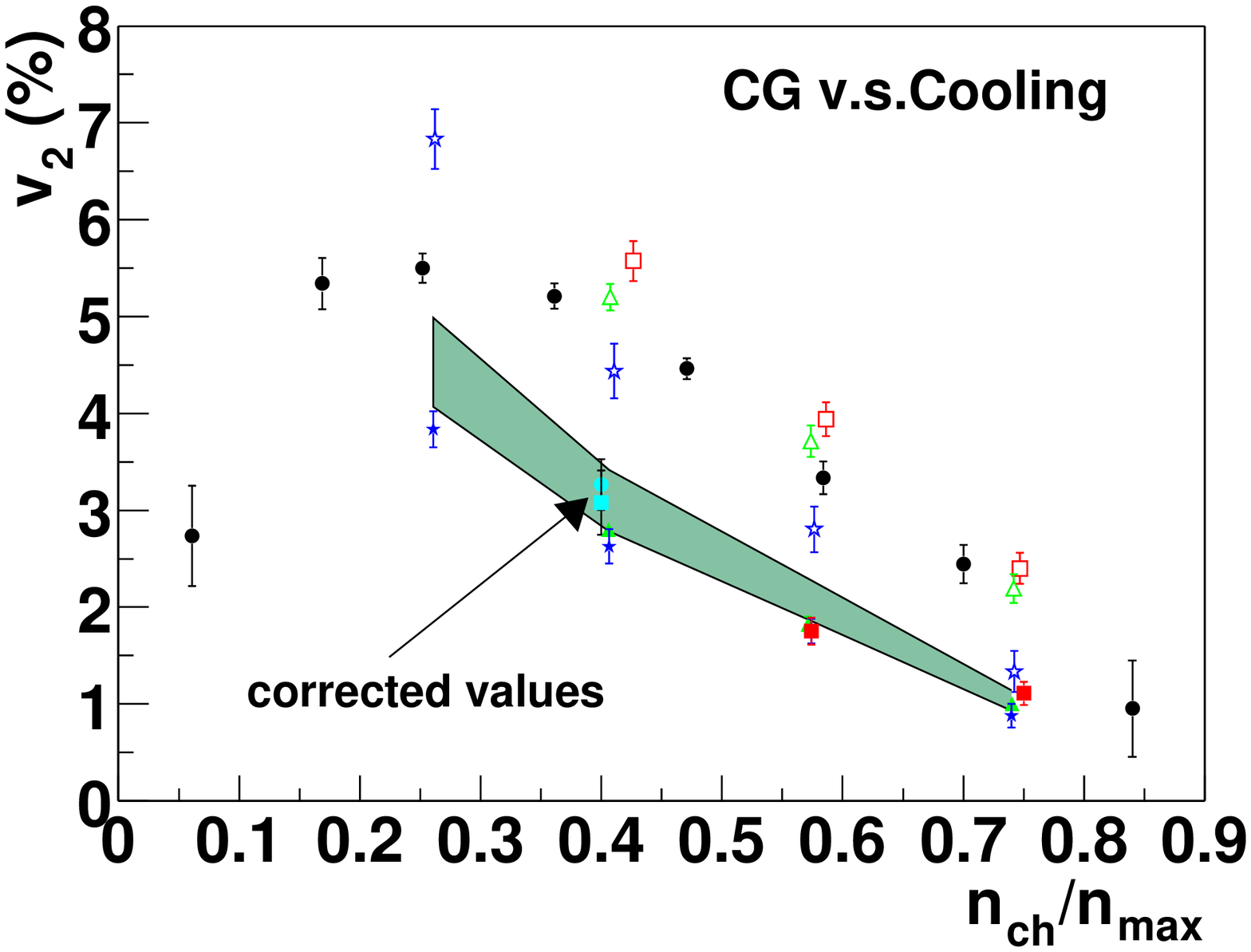,
        width=0.50\textwidth}}}
        \caption{a) A hydrodynamic fit to v2.  b) The Colored Glass fit.}
        \label{flow1}
\end{figure}

In off zero impact parameter heavy ion collisions, the matter which
overlaps has an asymmetry in density relative to the reaction plane.
This is shown in the left hand side of Fig. \ref{flow}.  Here
the reaction plane is along the x axis.  In the region of overlap there
is an $x-y$ asymmetry in the density of matter which overlaps.
This means that there will be an asymmetry in the spatial gradients which
will eventually transmute itself into an asymmetry in the momentum
space distribution of particles, as shown in the right hand side of
Fig. \ref{flow}.
This asymmetry is called elliptic
flow and is quantified by the variable $v_2$.  In all
attempts to theoretically describe this effect, one needs very strong
interactions among the quarks and gluons at very early times in the 
collision.\cite{flowth}.  In Fig. \ref{flow1}, two different theoretical 
descriptions are fit to the data by STAR and 
PHOBOS\cite{flowstar}-\cite{flowphobos}. On the left hand side, a 
hydrodynamical model is used.\cite{heinzkolb}  
It is roughly of the correct order of magnitude
and has roughly the correct shape to fit the data.  This was not
the case at lower energy.  On the right hand side are preliminary
fits assuming Color 
Glass.\cite{rajualex}  
Again it has roughly the correct shape and magnitude to describe
the data.  In the Color Glass, the interactions are very strong essentially
from $t = 0$, but unlike the hydrodynamic models it is field pressure
rather than particle pressure which converts the spatial anisotropy into
a momentum space-anisotropy.  

Probably the correct story for describing flow
is complicated and will involve both the Quark Gluon Plasma and the
Color Glass Condensate.  Either description requires that matter be produced
in the collisions and that it interacts strongly with itself.  In the limit
of single scatterings for the partons in the two nuclei, no flow is generated.

\subsection{What Do We Expect to Learn?}

\subsubsection{Does the Matter Equilibrate?}

\begin{figure}[htb]
    \centering
       \mbox{{\epsfig{figure=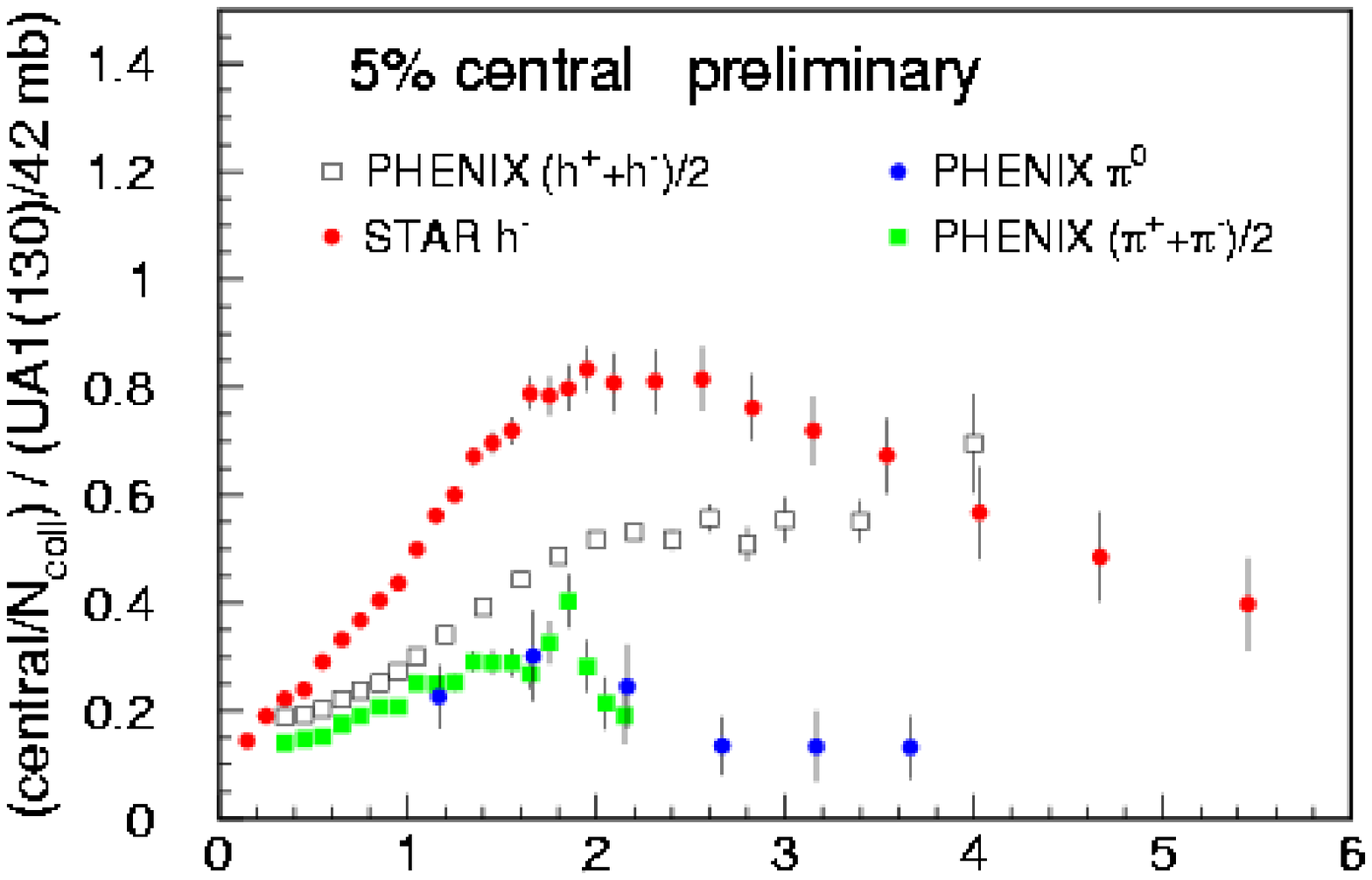,
        width=0.50\textwidth}}\quad
             {\epsfig{figure=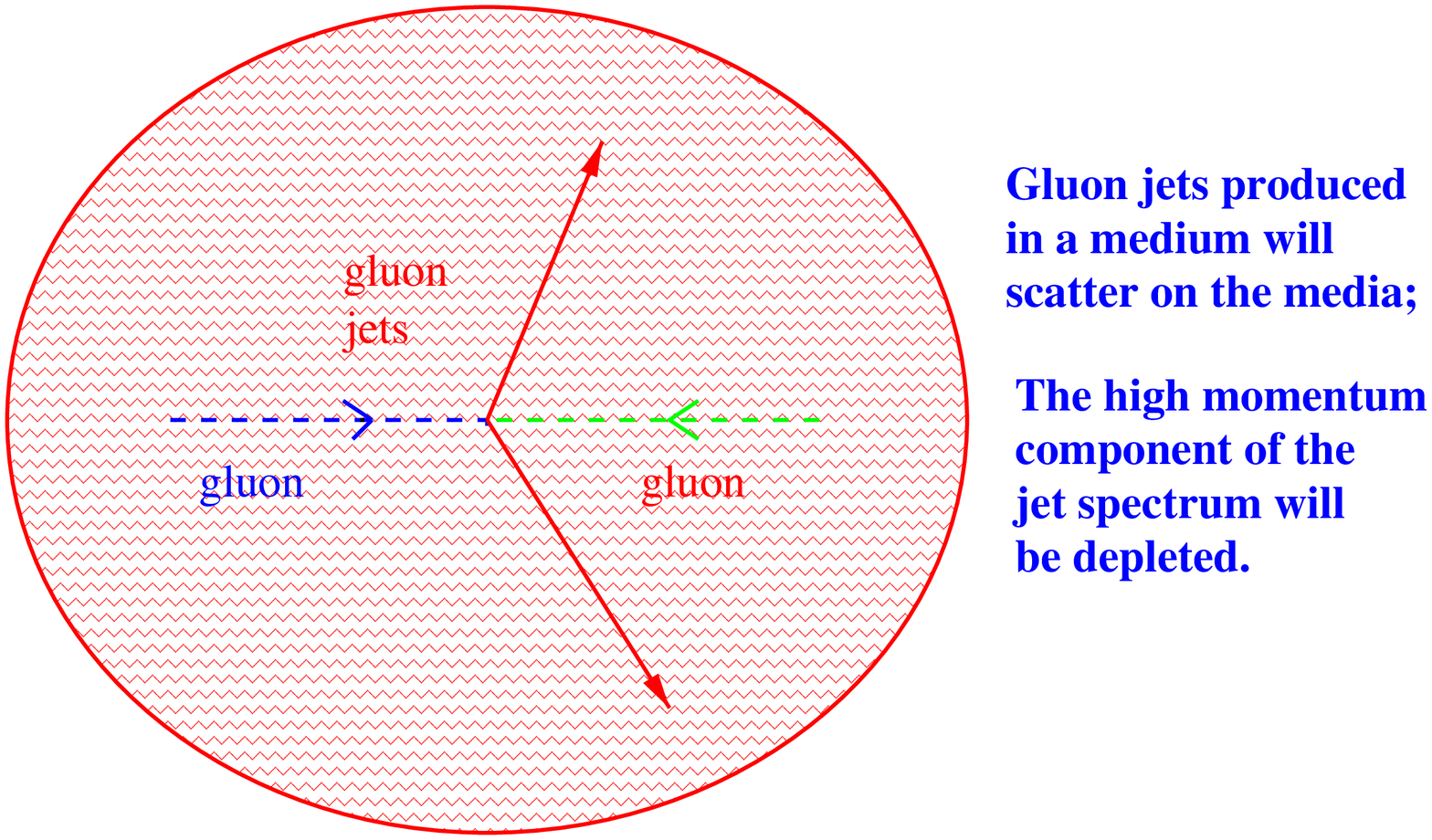,
        width=0.50\textwidth}}}
        \caption{a) The $p_T$ distribution of particles
scaled by the data from $pp$ collisions times the number of hard collisions
inside the nuclei.  b) A pair of jets is produced in a 
hard collision and they propagate through the surrounding matter.}
        \label{jets}
\end{figure}

One of the most interesting results from the RHIC experiments
is the so called ``jet quenching''.\cite{starjet}-\cite{phenixjet}.
In Fig. \ref{jets}a, the single particle hadron spectrum is scaled
by the same result in $pp$ collisions and scaled by the number of
collisions.  The number of collisions is the number of nucleon-nucleon 
interactions, which for central collisions should be almost all of the 
nucleons.  One is assuming hard scattering in computing this number,
so that a single nucleon can hard scatter a number of times as it penetrates 
the other nucleus.  The striking feature of this plot is that
the ratio does not approach one at large $p_T$.  This would be the value
if these particles arose from hard scattering which produced jets
of quarks and gluons, and the jets subsequently decayed.

The popular explanation for this is shown in Fig. \ref{jets}b.  Here a
pair of jets is produced in a gluon-gluon collision.  The jets are
immersed in a Quark Gluon Plasma, and rescatter as they poke through
the plasma.  This shifts the transverse momentum spectrum down, and the
ratio to $pp$ collisions, where there is no significant surrounding media,
is reduced.

The data, however suggestive, need to be improved before strong 
conclusions are drawn.  For example, there are large systematic
uncertainties in the $pp$ data which was measured in different detectors
and extrapolated to RHIC energy.  This will be resolved by measuring $pp$ 
collisions at RHIC.  There is in addition 
some significant uncertainty in the AA data which becomes smaller
in the ratio to $pp$ data when the data is measured in the same detector.
There are nuclear modifications of the gluon distribution function,
an effect which can be determined by measurements on $pA$ at RHIC.
The maximum transverse momentum is limited by the event sample size,
and the size will be greatly improved with this years run due to the
higher luminosity and longer run time.

(After these lectures were given, results were presented from the dA
experiments at RHIC.  The experimenters claim that the initial state 
effects are disfavored as an explanation for the jet quenching seen
in the Au-Au collisions.  For a summary of their conclusions, please look at 
the power point presentations on the RHIC homepage, 
or the submitted
papers.\cite{da}  It is the authors opinion that much more needs be done
before this conclusions can be firmly established.)

One of the reasons why jet quenching is so important for the RHIC
program is that it gives a good measure of the degree of thermalization
in the collisions.  If jets are strongly quenched by transverse momenta of
$4 ~GeV$, then because cross sections go like $1/E^2$ for quarks and gluons,
this would be strong evidence for thermalization at the lower energies
typical of the emitted particles.

One can look for evidence of thermalization directly from the measured $p_T$
distributions.  Here one can do a hydrodynamic computation and in so far as
it agrees with the results, one is encouraged to believe that there is
thermalization.  On the other hand, these distributions may have their
origin in the initial conditions for the collision, the Colored Glass.
In reality, one will have to understand the tradeoff between both effects.
The hydrodynamic models do a good job in describing
the data for $p_T \le 2~GeV$,  Here there is approximate $m_T$ scaling,
a characteristic feature of hydrodynamic computations.  This scaling
arises naturally because in hydrodynamic distribution are produced
by flowing matter which has a characteristic transverse flow 
velocity with a well defined local temperature.  Particles with the same
$m_T$ should have arisen from regions with the same transverse flow velocity
and temperature.

\begin{figure}[htb]
    \begin{center}
        \includegraphics[width=0.50\textwidth]{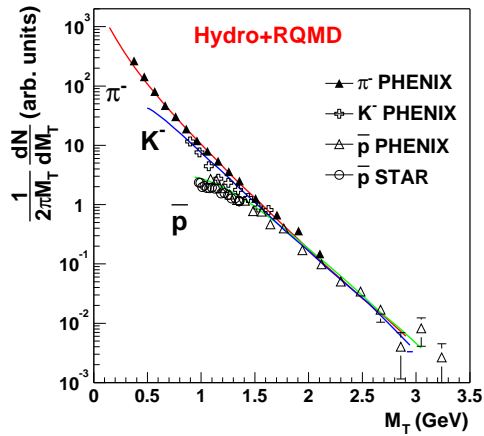}
        \caption{The hydrodynamical model fits to the $m_T$ spectra
for the PHENIX and STAR data.  
                            }\label{teaney}   
 \end{center}
\end{figure}

Hydrodynamical models  successfully describe the data on 
$m_T$\linebreak  distributions.\cite{shuryak}  In Fig. \ref{teaney}
the results of the simulation by Shuryak and Teaney are shown compared
to the STAR and PHENIX data.\cite{starjet}-\cite{phenixjet}  The shape of the
curve is a prediction of the hydrodynamic model, and is parameterized somewhat
by the nature of the equation of state.  Notice that the
STAR data include protons near threshold, and here the $m_T$ scaling 
breaks down.  The hydrodynamic fits get this dependence correctly,
and this is one of the best tests of this description.
The hydrodynamic models do less well on fits to the more peripheral collisions.
In general,
a good place to test the hydrodynamic models predictions is with massive
particles close to threshold, since here one deviates in a computable way
from the $m_T$ scaling curve, which is arguably determined from 
parameterizing the equation of state.

If one can successfully argue that there is thermalization in the RHIC
collisions, then the hydrodynamic computations become compelling.
One should remember that hydrodynamics requires an equation of state plus 
initial conditions, and these initial conditions are determined by Colored
Glass.  Presumably, a correct description will require the inclusion of both
types of effects.\cite{srivastava}

\subsubsection{Confinement and Chiral Symmetry Restoration}
We would like to know whether or not deconfinement has occurred in dense matter
or whether chiral symmetry restoration has changed particle masses.
\begin{figure}[htb]
    \begin{center}
        \epsfxsize = 3in
        \epsfbox[20 100 550 700]{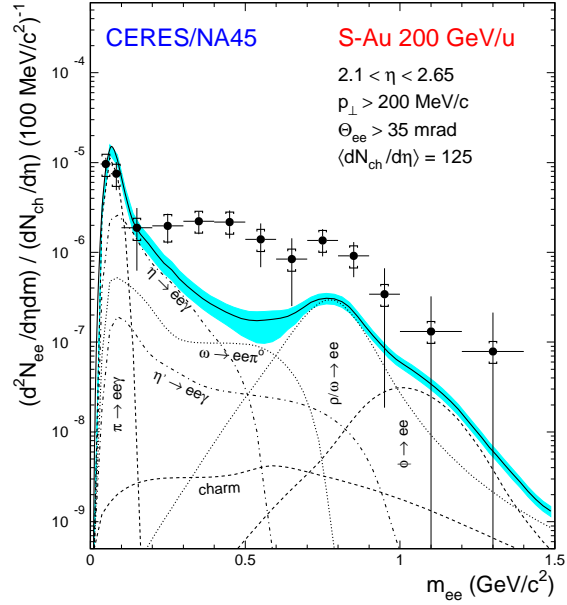}
        \caption{The CERES data on low mass electron-positron pairs.  
The expected 
contribution from ordinary hadrons is shown by the solid line. The data 
points are for the measured electron-positron pairs.   
                            }\label{ceres1}   
 \end{center}
\end{figure}


This can be studied in principle by measuring the spectrum of dileptons
emitted from the heavy ion collision.  These particles probe the interior
of the hot matter since electromagnetically interacting particles are
not significantly attenuated by the hadronic matter.  For electron-positron 
pairs, the mass distribution has been measured in the CERES experiment at 
CERN\cite{ceres}, and is shown in Fig. \ref{ceres1}.  Shown in the plot
is the distribution predicted from extrapolating from $pA$ collisions.
There should be a clear $\rho $ and $\phi $ peak, which has disappeared.
This has been interpreted as a resonance mass shift,\cite{brown},
enhanced $\eta^\prime$ production, \cite{kapusta} but is most probably
collisional broadening of the resonances in the matter
produced in the collisions.\cite{rapp}  Nevertheless, if one makes a plot
such as this and the energy density is very high and there are no resonances
at all, then this would be strong evidence that the matter is not hadronic, 
i. e. the hadrons have melted.  

The resolution in the CERES experiment is unpleasantly large, making it
difficult to unambiguously interpret the result.  Whether or not
such an experiment could be successfully run at RHIC, much less whether
the resolution could be improved, is the subject of much internal debate 
among the RHIC experimentalists.  

\subsubsection{Confinement and $J/\Psi$ Suppression}

\begin{figure}[htb]
    \centering
       \mbox{{\epsfig{figure=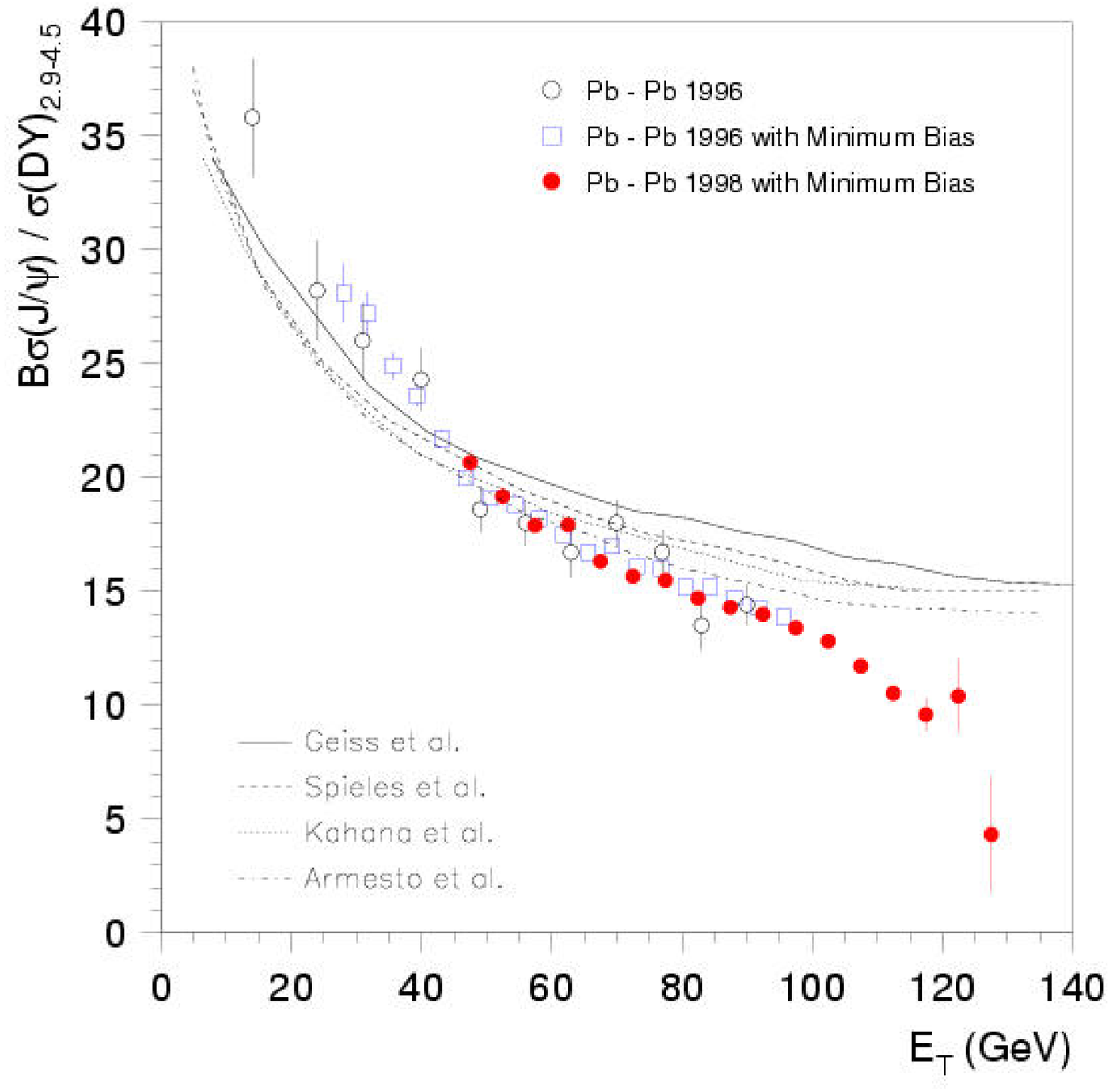,
        width=0.50\textwidth}}\quad
             {\epsfig{figure=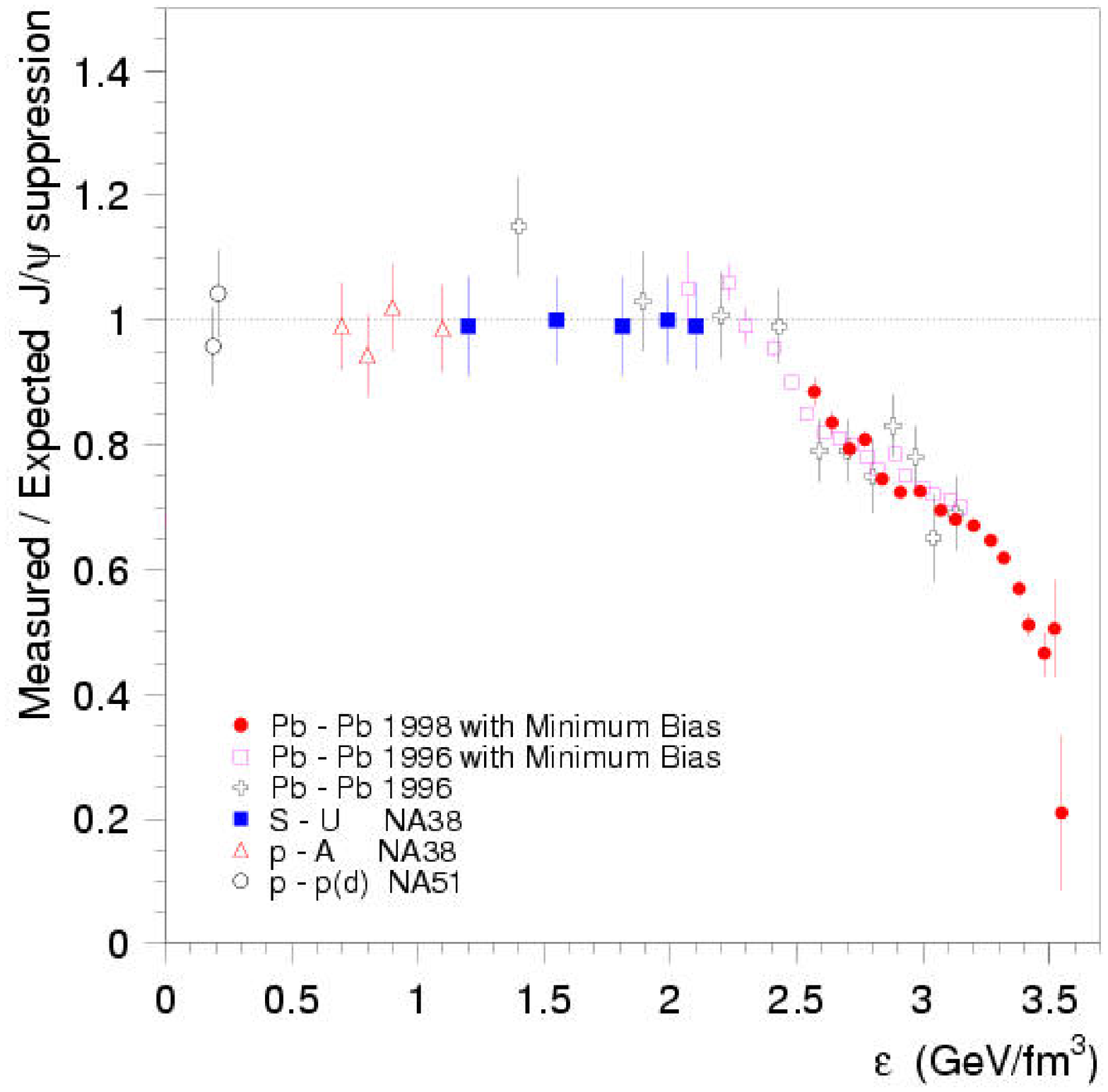,
        width=0.50\textwidth}}}
        \caption{a) 
The ratio of produced $J/\Psi$ pairs to Drell-Yan pairs as a function of 
transverse energy $E_T$ at CERN energy.  b) The measured
compared to the theoretically expected $J/\Psi$ suppression as a function
of the Bjorken energy density for various targets and projectiles.}
        \label{jpsi}
\end{figure}

In Fig. \ref{jpsi},  the NA(50) data for $J/\Psi$ production is 
shown.\cite{na50}
In the first figure, the ratio of $J/\Psi$ production cross section
to that of Drell-Yan is shown as a function of $E_T$,
the transverse energy, for the lead-lead 
collisions at CERN.  There is a clear suppression at large $E_T$ which
is greater than the hadronic absorption model calculations which are plotted
with the data.\cite{hadab}  In the next figure, the 
theoretically expected $J/\Psi$ suppression 
based on hadronic absorption models is compared
to that measured as a function of the Bjorken energy density
for various targets and projectiles.  There appears to be a sharp increase
in the amount of suppression for central lead-lead collisions.

Is this suppression due to Debye screening of the confinement potential
in a high density Quark Gluon Plasma?\cite{matsui}-\cite{blaizot}
Might it be due higher twists, enhanced rescattering, or changes in the gluon
distribution function?\cite{capella}-\cite{qiu}  Might the $J/\psi$
suppression be changed into an enhancement 
at RHIC energies and result from the 
recombination in the produced charm particles, many more of which are
produced at RHIC energy?\cite{rafelski}-\cite{gorenstein}

These various descriptions can be tested by using the    
capability at RHIC to do $pp$ and $pA$ as well as $AA$.  
Issues related to multiple scattering, higher twist effects, and changes
in the gluon distribution function can be explored.  
A direct measurement of open charm will
be important if final state recombination of the produced open charm
makes a significant amount of $J/\Psi$'s.

\subsubsection{The Lifetime and Size of the Matter Produced}

\begin{figure}[htb]
    \begin{center}
        \includegraphics[width=0.50\textwidth]{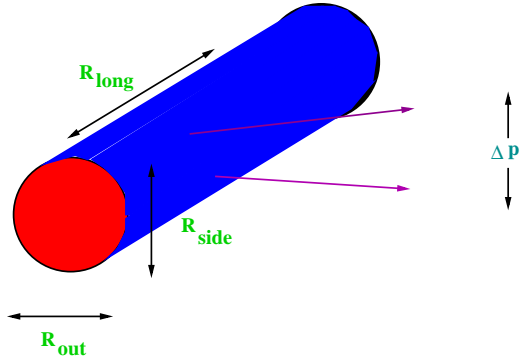}
        \caption{The various radii used for HBT pion interferometry.   
                            }\label{hbt1}   
 \end{center}
\end{figure}

The measurement of correlated pion pairs, the so called HBT pion
interferometry, can measure properties of the space-time volume from
which the hadronic matter emerges in heavy ion collisions.\cite{hbtreview}  
The quantities
$R_{long}, R_{side}$ and $R_{out}$ shown in Fig. \ref{hbt1} measure the
transverse size of the matter at decoupling and the decoupling time.
\begin{figure} [htb]
   \centering
       \mbox{{\epsfig{figure=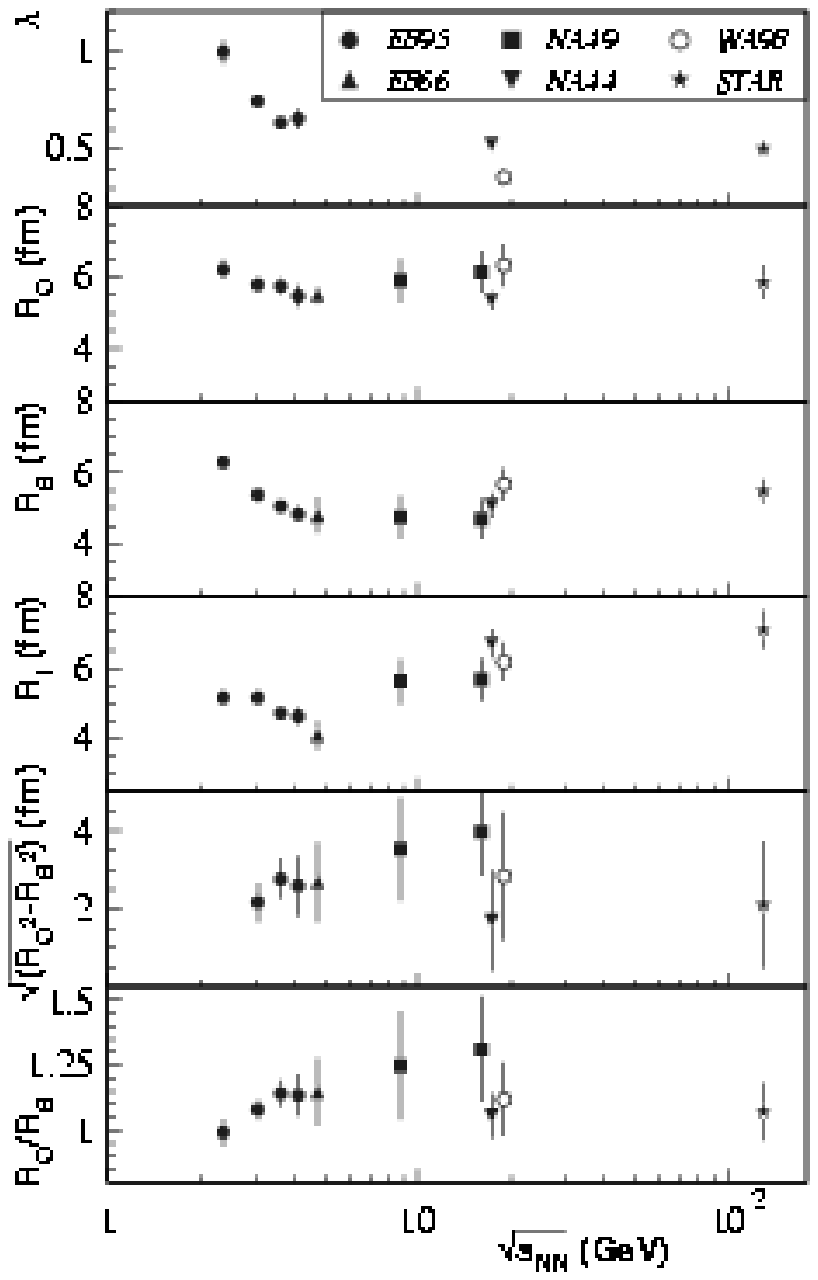,
        width=0.50\textwidth}}\quad
             {\epsfig{figure=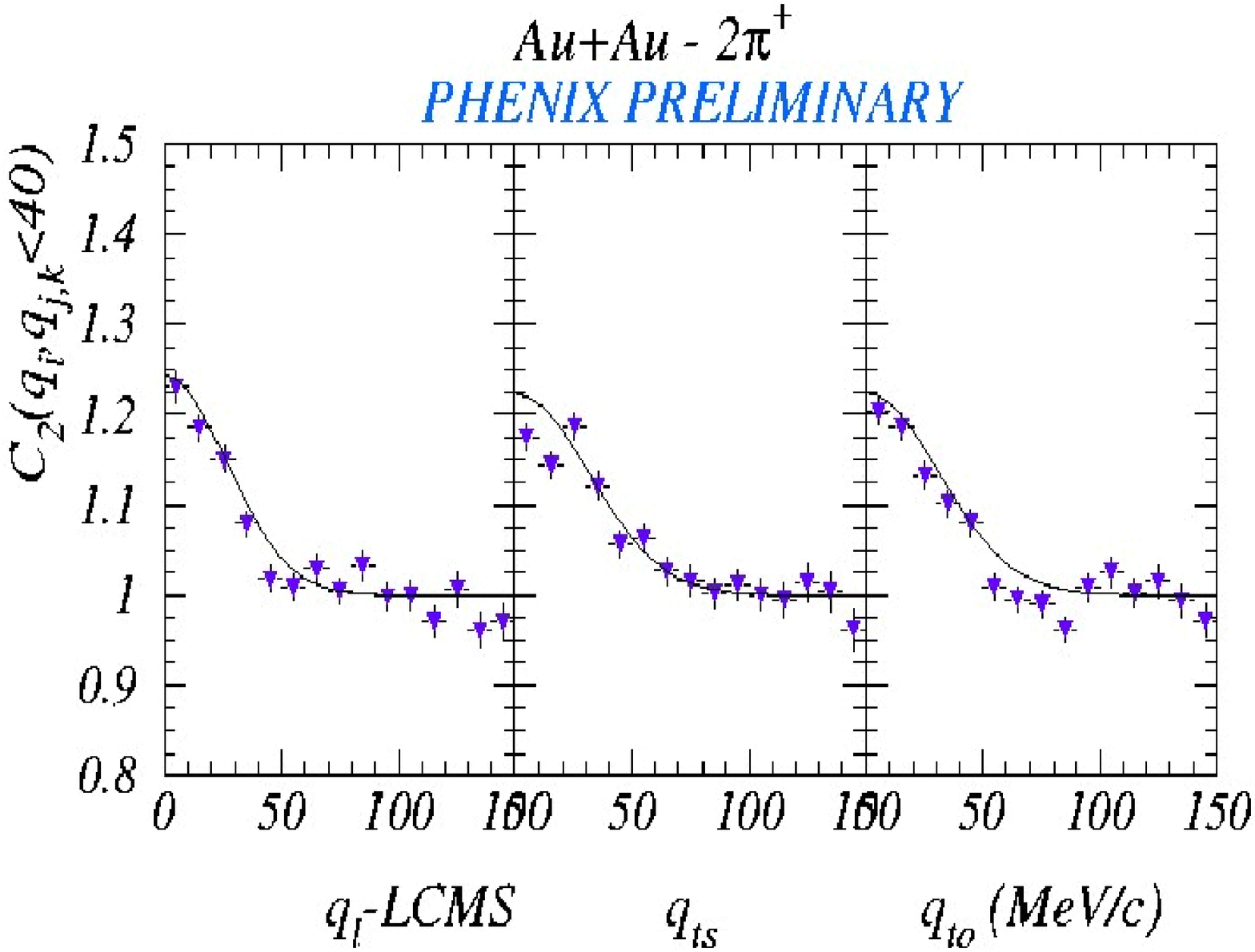,
        width=0.50\textwidth}}}
        \caption{a) The various HBT radii measured in heavy ion 
experiments including the new data from STAR.  b)
The correlation functions which determine the radii as a function of the
pair momenta measured in PHENIX.}
        \label{hbt2}
\end{figure}

In Fig. \ref{hbt2}, the data from STAR and PHENIX 
is shown.\cite{hbtstar}-\cite{hbtphenix}  There is only a 
weak dependence on energy, and the results seem to be more or less what
was observed at CERN energies.  This is a surprise, since a longer time
for decoupling is expected at RHIC.  Perhaps the most surprising result is
that $R_{out}/R_{side}$ is close to 1, where most theoretical expectations give
a value of about $R_{out}/R_{side} \sim 2$.\cite{heinzhbt}-\cite{soffhbt}
Perhaps this is due to greater than expected opacity of the emitting 
matter?   At this time, there is no consistent theoretical description
of the HBT data at RHIC.  
Is there something missing in our space-time picture?

\subsubsection{The Flavor Composition of the Quark Gluon Plasma}

The first signal proposed for the existence of a Quark Gluon Plasma in heavy
ion collisions was enhanced strangeness production.\cite{muller}
This has lead to a comprehensive program in heavy ion collisions to measure 
\begin{figure} [htb]
   \centering
       \mbox{{\epsfig{figure=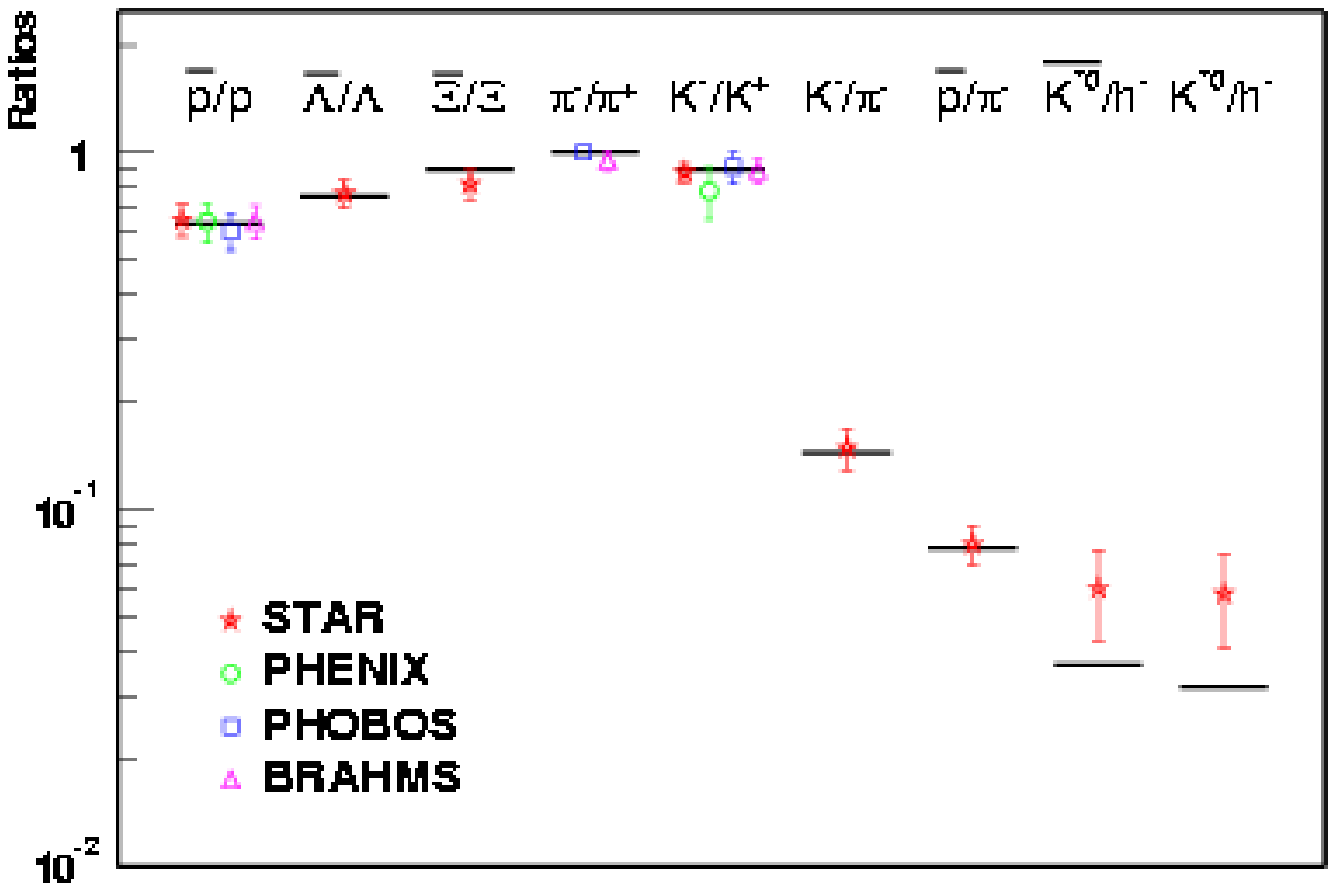,
        width=0.50\textwidth}}\quad
             {\epsfig{figure=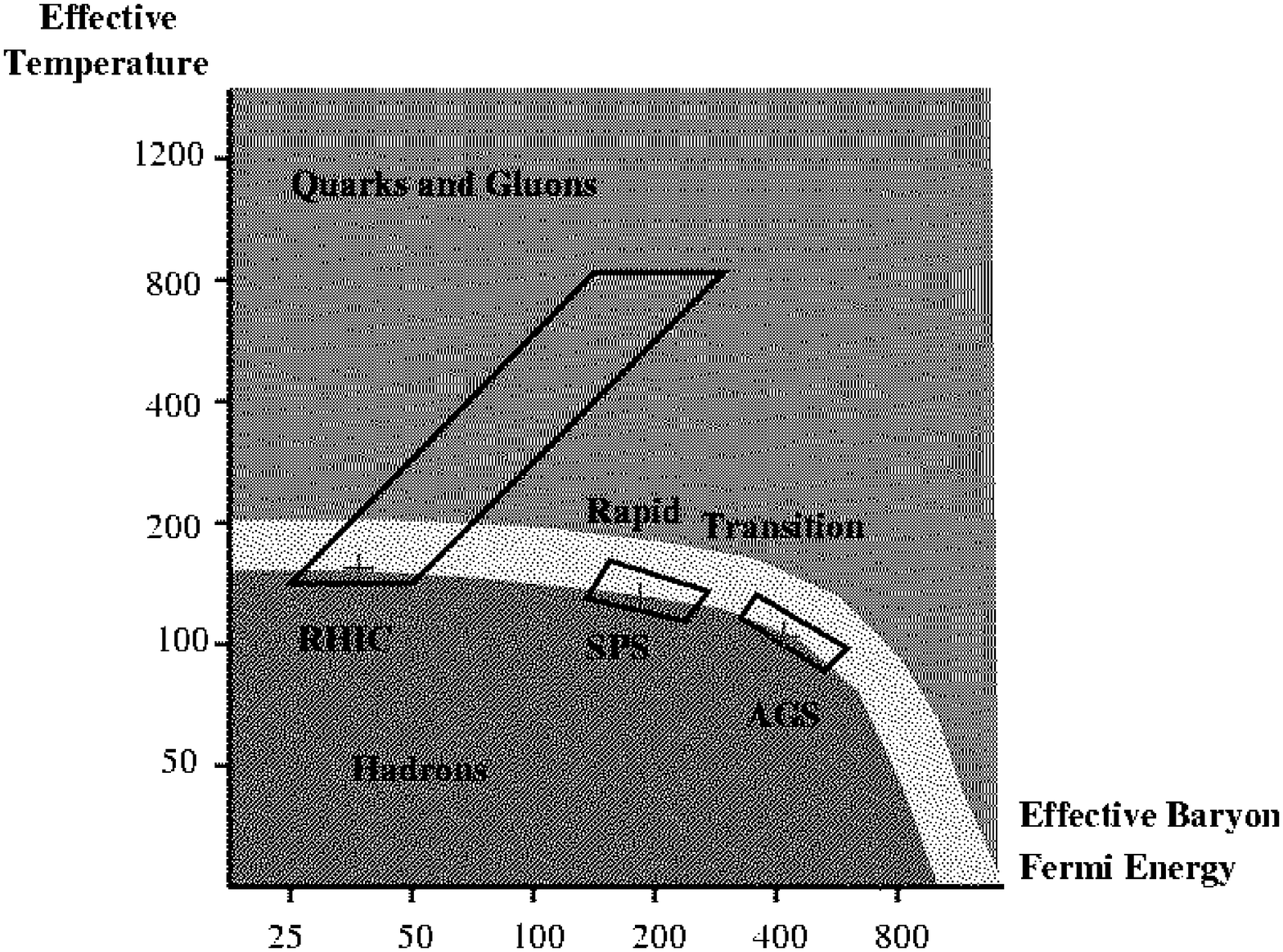,
        width=0.50\textwidth }}}
        \caption{a) Various ratios of particle abundances and
the RHIC data.  The lines are the predictions of a thermal model.  b) 
The temperature vs baryon chemical potential for a thermal model
which is fit to data at various energies.  The dashed line is a 
hypothetical phase boundary between a Quark Gluon Plasma and a hadronic gas.}
        \label{chem}
\end{figure}
the ratios of abundances of various flavors of particles.\cite{nuxu}.
In Fig. \ref{chem}a, the ratios of flavor abundances is compared
to a thermal model for the particle 
abundances.\cite{cleymans} - \cite{gorenstein1}  The fit is quite good.
In Fig. \ref{chem}b, the temperature and baryon chemical potential 
extracted from these fits is shown for experiments at various energies
and with various types of nuclei.  It seems to agree reasonably well
with what might be expected for a phase boundary between hadronic
matter and a Quark Gluon Plasma.

This would appear to be a compelling case for the production of a 
Quark Gluon Plasma.  The problem is that the fits done for heavy ions to
particle abundances work even better in $e^+e^-$ collisions.  One definitely
expects no Quark Gluon Plasma in $e^+e^-$ collisions.  There is a deep
theoretical question to be understood here:  How can thermal models
work so well for non-thermal systems?  Is there some simple saturation of
phase space?  The thermal model description can eventually be made compelling
for heavy ion collisions once the degree of thermalization in
these collisions is understood.

\section{Lecture 4:  The Physics of the Color Glass Condensate}

In this lecture, I discuss some of the implications of the Color 
Glass Condensate.  I begin by developing in a little more detail the 
solutions for the fields of the Color Glass Condensate for a single hadron.
I then discuss issues related to unitarity for electromagnetic probes
of hadrons.
I later argue that 
Froissart bound saturation for the total cross section in hadron-hadron
scattering also arises naturally in the context of the
Color Glass Condensate.  Next, I show that a new scale appears
in the gluon structure function and which corresponds to 
a new kind of Geometrical 
Scaling of high energy deep inelastic scattering.  This scaling implies
that there is a form of matter intermediate between the
highly coherent Color Glass Condensate and the incoherent parton
densities of perturbative QCD.  Finally, I discuss
the renormalization group 
and its implications for the high energy limit.

\subsection{Formal Development: Light Cone Quantization}

\begin{figure}[ht]
    \begin{center}
        \includegraphics[width=0.50\textwidth]{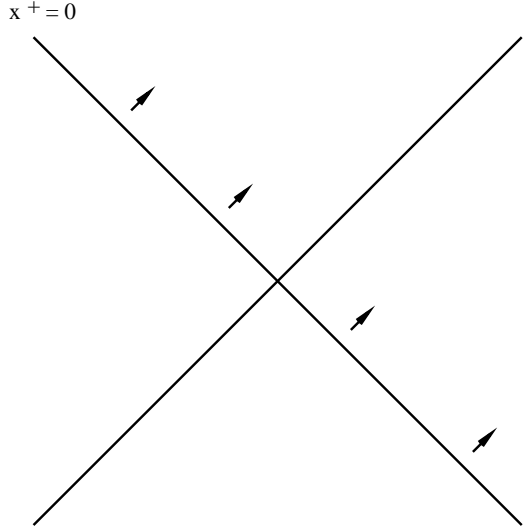}
        \caption{The initial value problem in lightcone coordinates.
The light cone is shown and propagation from initial conditions
at $x^+ = 0$ is shown. }\label{lightconein}
    \end{center}
\end{figure}

In light cone coordinates, the initial value problem is formulated
along the surface $x^+ = 0$, and propagation is in terms of the light
cone time $x^+$.  This is shown in Fig. \ref{lightconein}.

Quantization of fields in light cone coordinates is most easily 
illustrated for the Klein-Gordon field.  The equation of motion is
\be
	(p^2 - M^2) \phi = 0
\ee
or 
\be
	p^- \phi = {{p_T^2 + M^2} \over {2p^+}} \phi
\ee
Since $p^- = i \partial / \partial^+$, the first quantized Hamiltonian 
corresponding to this equation is
\be
	p^- = {{p_T^2 + M^2} \over {2p^+}}
\ee

To second quantize this system, we consider the action
\be
	S = \int d^4x {1 \over 2} \left( (\partial \phi )^2 - 
M^2 \phi^2 \right)
\ee
The canonical momentum is
\be
	\Pi (x^-,x_T) = {{\delta S} \over {\delta \partial_+ \phi }}
= \partial^+ \phi = {\partial \over {\partial x^-}} \phi
\ee
Note that the momentum $\Pi$ is on the same equal time
surface as is $\phi$, and therefore it is not a variable independent of
$\phi$.  The momentum and coordinate are constrained, and therefore
the quantization is subtle.   

If we postulate the equal-time commutation relations
\be
	[\Pi(x^-,x_T), \phi(y^-,y_T)] = {{-i} \over 2}~ \delta^{(3)} (x-y)
\ee
we see that
\be 	\partial_-^x [\phi(x),\phi(y) ] = {{-i} \over 2} \delta^{(3)}
(x-y)
\ee
or that
\be
	[\phi(x), \phi(y) ] = {{-i} \over 2} ~\epsilon (x^- - y^-) 
\delta^{(2)} (x_T - y_T)
\ee
One can check that these commutation relations generate
the correct equations of motion for the action above.  This leads to the 
creation and annihilation operator basis for the field $\phi$ of
\be
	\phi (x) = \int {{d^3p} \over {(2\pi)^3 2p^+}} \theta (p^+) 
\left( e^{-ipx} a(p) + e^{ipx} a^\dagger (p) \right)
\ee
where
\be
	[a^b_i(p) ,a^c_j(q) ] = 2 p^+ \delta^{bc} \delta_{ij} \delta^{(3)}
(p-q)
\ee
Note that  on the light one, only positive $p^+$ particles propagate.  The 
vacuum has zero $p^+$, and since momentum is conserved, is trivial and 
has no particles in it.  (This is true for the vacuum built to any
order in perturbation theory.  In fact the light cone limit is very
subtle, and when one is careful to properly treat modes that have $p^+ =0$,
these modes can generate non-perturbative condensates for the vacuum.)

To quantize QCD, we work in light cone gauge, $A^+_a = 0$. The equation
of motion
\be
	D_\mu F^{\mu +} = - D_i F^{i +} + D^+ F^{-+} = 0
\ee
so that
\be
	A^- = {1 \over \partial^{+2}} D^i \partial^+ A^i 
\ee
The transverse degrees of freedom are quantized as
\be
	A^i_a = \int {{d^3p} \over {(2\pi)^3 2p^+}} \left( e^{-ipx} a^a_i (p)
+ e^{ipx} a_i^{a\dagger}(p) \right)
\ee

If we want to compute the gluon content of the hadron, we compute
\be
	{{2p^+} \over {(2\pi)^3}} = <h \mid a^\dagger (p) a(p) \mid h >
= {{2p^+} \over {(2\pi)^3}} <h \mid A^{ia}(p) A_{ia} (p) \mid h >
\ee
which we recognize as $2p^+/(2\pi)^3$ times the propagator
$G^{ii}_{aa}(p,p; x^+-y^+ \rightarrow 0)$

\subsection{Formal Development: Solving the McLerran-Venugopalan Model}

The McLerran-Venugopalan (MV) model involves 
the computation of the classical fields
due to a lightcone current, and then averaging with a Gaussian
weight over the external current strength.  It is the simplest model
with the physics of saturation built in.  Instead of working in light cone 
gauge, it is simplest to first solve the classical equations of
motion in the gauge $A^- = 0$, and then to gauge rotate the result back
to lightcone gauge.  (The action for a Gaussian source can be written
in a gauge invariant way because all values of the sources
are integrated over with a gauge invariant measure.)

The equations of motion in $\overline A^- = 0$ gauge are
\be
	D_\mu F^{\mu \nu} = \delta^{\nu + } \rho(x^-,x_T)
\ee
(We overline fields to indicate these are the fields in this gauge which must
be rotated back to the light cone gauge.  Fields in the light cone gauge will
not be overlined.)
These equations can be solved by the fields
\be
	\overline A^i = 0
\ee
and
\be
	- \nabla^2_T \overline A^+ = \overline \rho
\ee
Note that
\be
	\overline \rho = U^\dagger \rho U
\ee
where $U$ is the gauge rotation between this gauge and light cone gauge.

Since 
\be
	\overline A^\mu = U^\dagger A^\mu U + {i \over g} U^\dagger 
\partial^\mu U
\ee
we have that
\be
	\overline A^+ = {i \over g} U^\dagger (\partial^+ U)
\ee
Upon defining 
\be
	\alpha = \overline A^+ = { 1 \over {-\nabla_T^2}} \overline \rho
\ee
we see that we can explicitly determine
\be
	U^\dagger (x) = P exp \left( ig \int_{-\infty}^{x^-} dz^- 
\alpha(z^-,x_T) \right)
\ee
The fields in $A^+ = 0$ gauge are therefore 
\be
	A^+ = A^- = 0
\ee
and
\be
	A^i = {i \over g} U \nabla^i U^\dagger
\ee
If we choose $x^-$ to be outside the range of support of $\rho$,
then these fields are of the simple form
\be
	A^i = \theta (x^-) V \nabla^i V^\dagger
\ee
where
\be
	V^\dagger (x) = Pexp\left( ig \int_{-\infty}^{\infty} dz^-
\alpha(z^-,x_T) \right)
\ee
\subsection{The Gluon Distribution Function in the MV Model}

We can now use the classical fields we determined above to compute the
gluon distributions function.  We have that
\be
	{{dN}\over {d^3k}} = {{2k^+} \over {(2\pi)^3}} <A^i_a (k,x^+) 
A^i_a(-k,x^+)>
\ee
where the $<O>$ notation means to average $O$ over all values of
sources with a Gaussian weight.  This averaging is straightforward to do
for the gluon distribution function with the result 
that \cite{JKLW97}
(in coordinate space for both $x^-$ and $y^-$ greater than zero)
\be
	<A^i_a(x) A^i_a(y)> = {{N_c^2-1} \over {\pi \alpha N_c}}
{{1 - e^{-x_T^2 Q_s^2 ln(x_T^2 \Lambda^2_{QCD})/4}} \over x_T^2}
\ee
The saturation momentum is \cite{AM2}-\cite{PI}
\be
	Q_s^2= 2\pi N_c \alpha_s^2 \int dx^- \mu^2 (x^-) \sim \alpha_s^2
{{Charge^2} \over {area \times (N_c^2 -1)}}
\ee
This equation is true only for $x_T << 1/\Lambda_{QCD}$, and also assumes
that the scale of charge neutralization is the confinement scale.  This 
neutralization scale  becomes modified to $Q_s$ in a more first principles
computation and the structure of the distribution function is 
modified for $1/\Lambda_{QCD} \ge x_T \ge 1/Q_s$.
Note that the integral over $x^-$ in the definition of the 
saturation momentum can be converted to an integral over space time
rapidity and that the charge being computes is the total at all rapidities
greater than that of the scale of interest.  One can use the DGLAP 
evolution equations to relate this directly to the gluon density.

\begin{figure}[ht]
    \begin{center}
        \includegraphics[width=0.50\textwidth]{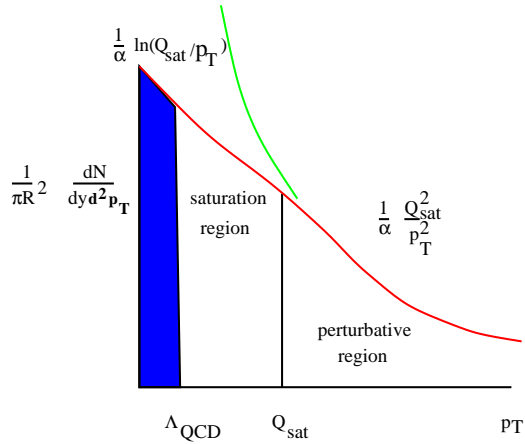}
        \caption{The gluon distribution function as
computed in the Mclerran-Venugopalan model. }\label{mvglue}
    \end{center}
\end{figure}

The gluon distribution can now be computed in momentum space using the above 
formulae.  One finds that at large $p_T$,
it goes as $Q_s^2/\alpha_s p_T^2$, and at small $p_T$, it goes as
$ln(Q_s^2/p_T^2)/\alpha_s$.  The gluon distribution is shown in
Fig. \ref{mvglue}
Note that the omnipresent factor  of $1/\alpha_s$ arises from the
strong gluon fields and is typical of condensation phenomena.

At large $p_T$, the behaviour as $Q_s^2/p_T$2 is typical of bremstrahlung
from a number of independent sources characterized by the gluon
distribution function.  At small $p_T$, these sources add together
coherently, and the monopole nature of the fields strength is canceled.
This is because at these $p_T$ scales, the typical separation between
sources is less than the scale at which the field is measured.

We expect that the saturation momentum will grow quickly with $1/x$,
perhaps a power of $1/x$.  The tail of the distribution in $p_T$ therefore
grows rapidly with $1/x$, but below $Q_s$, the distribution is slowly varying.
This behaviour solves the unitarity problem associated
with the energy dependence of distribution functions.  If we measure the
number of gluon below some resolution scale $Q^2$,
\be
	xG(x,Q^2) \sim \int_0^{Q^2} d^2p_T {{dN} \over {d^2p_T dy}}
\ee
this goes as $\pi R^2 Q^2$ for $Q \le Q_s$ and $\pi R^2 Q_s^2$ for
$Q \ge Q_s$.  So at fixed Q at very small x, it will always be true that
$Q \le Q_s$ and cross sections will grow geometrically.  In the MV model,
$Q_s^2 \sim R$, and therefore at large $Q \ge Q_s$, the cross section grows
rapidly and scales as the volume of the system.  (In fact at very small x
it turns out that the renormalization group equations give a $Q_s$ independent
of $R$, although there is still a rapid increase with energy.)

\subsection{Hadronic Cross Sections and Froissart Bound Saturation}

The Froissart bound is that a total hadronic cross section must
satisfy \cite{Froissart}-\cite{Martin}
\be
	\sigma \le Cons \times ln^2(E)
\ee
at high energy $E$.  This may be simply understood in the language
of the Color Glass Condensate.\cite{fbiim}-\cite{kw}  
Suppose that the saturation momentum
grows like an exponential in rapidity at small x $y \sim ln(1/x)$.
let us also assume that the saturation momentum is a function of rapidity 
and that it factorizes into an impact parameter dependent piece and a $y$
dependent piece,
\be
	Q_s^2(y,b) = Q_s^2(y,0) F(b)
\ee
Let us assume that 
\be
	Q_s^2(y,0) \sim e^{\kappa y}
\ee
At large $b$, we expect that $F(b) \sim e^{-2 m_\pi b}$, since an
isosinglet quantity like the gluon distribution function's large distance 
effect should be controlled by two pion exchange.

Now if we measure a cross section at some value of $Q^2$, then it better
be true that the target becomes dark at an impact parameter which satisfies
\be
	Q^2 = Q^2_s(y,b)
\ee
or that the impact parameter where this occurs satisfies
\be
	b \sim \kappa y/2m_\pi
\ee
Therefore the cross section saturates the Froissart bound
\be
	\sigma \sim y^2
\ee

To fill in the details of this argument involves much work and the interested
student is invited to explore the literature where these issues are discussed,
and are continuing to be argued, in the literature.

\subsection{Geometrical Scaling} 

Geometric scaling is the condition that the structure functions for 
quarks and gluons are functions only of the dimensionless ration
$Q^2/Q_s^2$, up to an overall factor which carries the overall 
dimension.\cite{GBW99}-\cite{KIIM}
For deep inelastic scattering, it is the requirement
\be
	\sigma_{\gamma^*p} \sim F_2(x,Q^2)/Q^2 \sim G(Q^2/Q_s^2)
\ee
This condition is obvious when $Q^2 << Q_s^2$, but a surprising result is that
it is also true up to $Q^2 \le Q_s^4/\Lambda^2_{QCD}$.  This weaker
bound can take one to quite high values of $Q^2$ at small values of $x$.

\begin{figure}[ht]
    \begin{center}
        \includegraphics[width=0.50\textwidth]{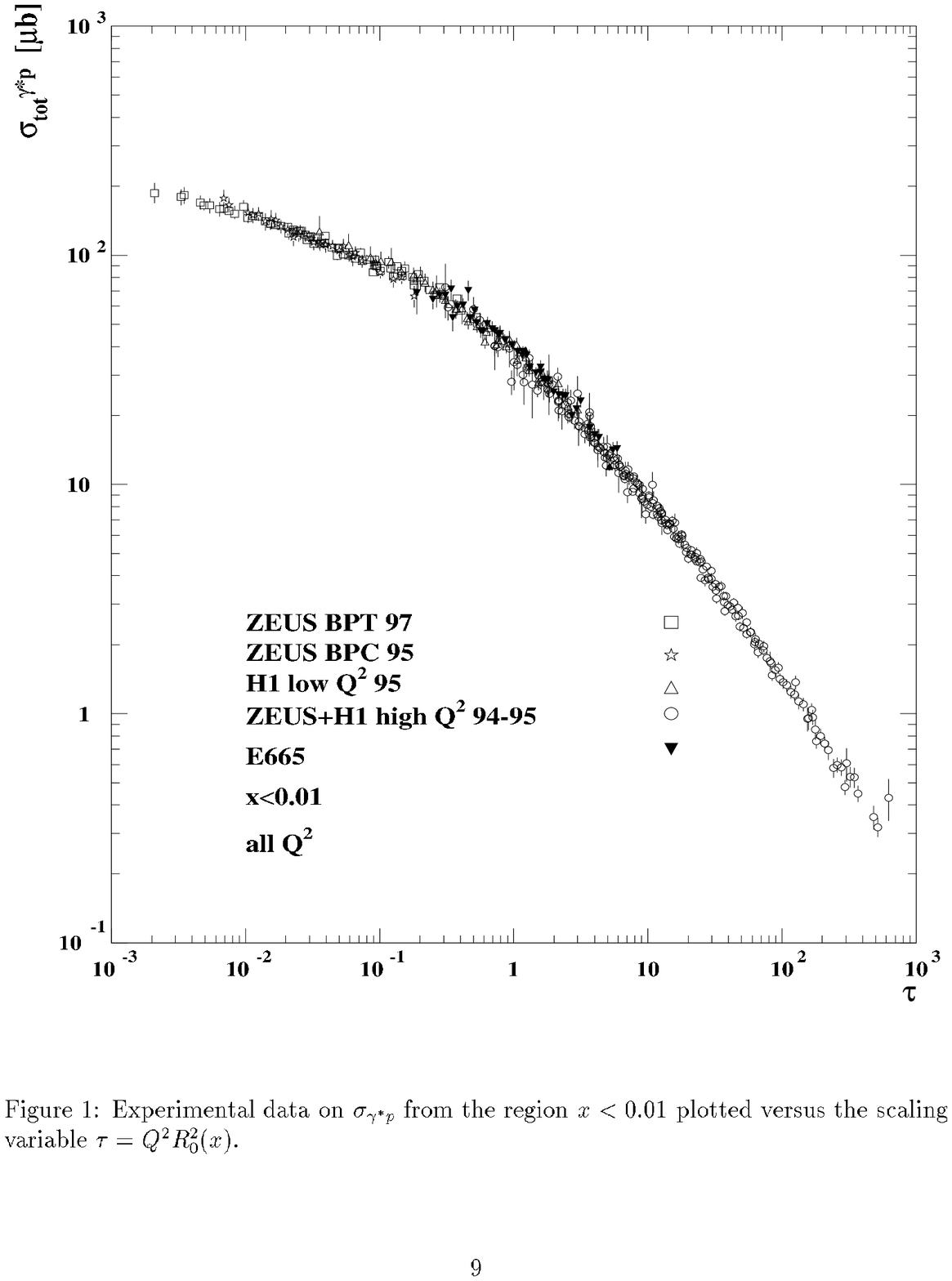}
        \caption{$\sigma_{\gamma^*p}$ as a function of the scaling variable
$\tau = Q^2/Q_s^2$. }\label{geom}
    \end{center}
\end{figure}

The worlds data at $x \le 10^{-2}$ is shown in Fig. \ref{geom}
as a function of $\tau = Q^2/Q_s^2$
It seems to scale in terms of the saturation momentum.

To understand how geometrical scaling can arise, consider the simple example
of the unintegrated 
gluon distribution function in the double logarithm approximation.
Here
\be
	F(x,Q^2) = \Lambda^2_{QCD} e^{\sqrt{Aln(1/x)ln(Q^2/\Lambda_{QCD}^2)}}
\ee
The gluon distribution function is the number of gluons per unit area
with momentum less than $Q^2$.  When $Q^2 \sim Q_{sat}^2$, this density
is of the order of $Q_{sat}^2$, as we saw in previous sections.  This gives
us an equation for the saturation momentum
\be
	Q_{sat}^2 = \Lambda_{QCD}^2 e^{Aln(1/x)}
\ee
This predicts a power law dependence upon the $x$ in agreement with
phenomenology.  This has a relatively large correction due to running
the coupling constant, but in fact one can compute the saturation
momentum's dependence on $x$ in a systematic way and the result, remarkably,
agrees with phenomenology.

Now if go back and express $\Lambda^2_{QCD}$ in terms of $Q_{sat}^2$,
and require that $Q^2 << Q_{sat}^4/\Lambda^2_{QCD}$, we find that
\be
	F(x,Q^2) \sim (Q^2/Q_{sat}^2)^{1 /2}
\ee
The gluon distribution function acquires an anomalous dimension of
$1/2$.  Again this can be done beyond the double logarithmic approximation,
and one can find the above arguments go through save that the anomalous 
dimension is a little changed from $1/2$, and that the power law dependence 
of the saturation momentum is changed.  The interested student is 
referred to the original literature to see this fully developed.

\begin{figure}[ht]
    \begin{center}
        \includegraphics[width=0.50\textwidth]{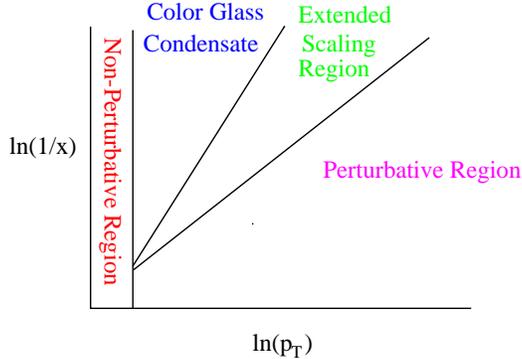}
        \caption{The kinematic regions in the $ln(1/x) - ln(Q^2)$
plane which correspond to Extended Scaling and to the Color 
Glass Condensate }\label{extscaling}
    \end{center}
\end{figure}

This result has the consequence that the effects of saturation extend far
beyond the region of momentum where there is a Color Glass Condensate
$Q^2 \le Q_{sat}^2$, into a new region 
$Q_{sat}^2 \le Q^2 \le Q_{sat}^4/\Lambda^2_{QCD}$.  In this new 
extended scaling region, distribution functions have pure power law
behaviour reminiscent of critical phenomena in condensed matter systems.
This region has been called the Extended Scaling Region and sometime the
Quantum Colored Fluid.  A diagram which shows the various
kinematic regions is shown in Fig. \ref{extscaling}.

\subsection{The Renormalization Group}

The result of our considerations for an effective action for the 
Color Glass Condensate gave  path integral representation of the form
\be
	\int [dA] [d\rho] exp\left(iS[A,\rho] - F[\rho] \right) 
\ee
In this equation $S$ is the action for the gluon fields in the presence of a 
light cone current described by the charge density $\rho$.  (To define
it properly one must provide a manifestly gauge invariant action.)
Once one solves for the fields in terms of $\rho$, then one is required
to average over the source with a weight function $F[\rho]$ which in the
Mclerran-Venugopalan model is taken as a Gaussian.  Implicit in the path
integral is a longitudinal momentum cutoff.  The fields have
momenta below this cutoff, and the effect of integrating out the
fields above the cutoff is included in the source $\rho$ and the integration
over various values of $\rho$.

The question arises:  How does one determine $F[\rho]$?  It turns out that $F$
is determined by renormalization group equations generated by varying the
longitudinal momentum cutoff.\cite{JKMW97}-\cite{JKLW97},\cite{PI},
\cite{Balitsky1996}-\cite{AM01}
The reason why the renormalization group 
treatment is essential follows from trying to solve for physical quantities,
such as the gluon distribution function within the CGC approach.  In lowest
order one computes the classical field associated with the source 
$\rho$, inserts it into an expression for the operator of interest,
and then averages over $\rho$.  The lowest order 
corrections to this involve Gaussian fluctuations around this classical
solution.  If there is some scale associated with the process of physical
interest, say $p^+$, one finds that the first order corrections
are of order $\alpha_s ln(\Lambda^+ /p^+)$, where $\Lambda^+$ is the
longitudinal momentum cutoff.  The coupling constant $\alpha$ is small
because we evaluate it at the saturation momentum scale.  The quantum 
corrections to the lowest order result are therefore small so long as
\be
	e^{-c/\alpha_s} \Lambda^+ << p^+ \le \Lambda^+
\ee

In order to go to lower momenta, it is easiest to change the
longitudinal momentum cutoff to a smaller value.  To do this, we 
have to integrate out the degrees of freedom between the old longitudinal 
momentum cutoff scale and the new one.  This can be done in 
Gaussian approximation since the coupling is weak, and so long
as the ratio of the various cutoff scales satisfies 
$\alpha_s ln(\Lambda^+/\Lambda^{+\prime}) << 1$.  It turns out that this
integration does not change the action for the interaction of the gluon fields
with the source.  All that changes is the weight function for integration
over the source fields.  If we let 
\be
	dy = ln(\Lambda^+/\Lambda^{+\prime})
\ee
the renormalization group equation becomes
\be
	{d \over {dy}} e^{-F[\rho ]} = -H(\rho, d/d\rho) e^{-F[\rho]}
\ee
It turns out that $H$ is second order in $d/d\rho$, real, and positive 
semidefinite.  Therefore $H$ can be interpreted as a Hamiltonian for
a $2+1$ dimensional quantum system. 

The Hamiltonian above has an unusual property.  If there was a potential
for the Hamiltonian $H$ with a unique minimum, then the at large times
the solution of the above equation would tend towards the ground state.
In the Color Glass Condensate $H$, there is in fact zero potential.
The system never tends to the ground state, and there is
quantum diffusion.  To see how this works consider a 1 dimensional example.
\be
	{d \over dy} Z = {{-p^2} \over 2} Z
\ee
This has the solution
\be
	Z = {1 \over \sqrt{2\pi y}} exp\left(-x^2/2y \right)
\ee
As the Euclidean time $y$ increases, the wavefunction spreads, corresponding
to diffusion.  This is unlike the situation where there is a potential
\be
	{d \over {dy}} Z = \left( {{-p^2} \over 2} -V(x) \right) Z
\ee
Here as we evolve in time, the coordinate $x$ settles into the minimum of
$V$, and has small excursions around it.  The solution for $Z$ becomes time
independent. 

The consequences of this simple observation are enormous.  For 
the case of diffusion, physical quantities are never independent of
rapidity, even at the smallest values of $x$.  The non-triviality of the small
$x$ limit is a consequence of the lack of a potential in the
renormalization group evolution equation!
                                       
The interested reader is referred to the growing literature on this subject
for details.  Suffice it to say that one can use the renormalization group
equation above, and the explicit form for $H$ which has been computed to 
reproduce all known renormalization group equations. The explicit form of the 
equations exists, and various approximate solutions have been constructed.
The picture which results agrees with the phenomenology of small x physics.
Understanding and solving these equations provides a rich area for future
research.

\section{Acknowledgements}

I grateful acknowledge conversations with Dima Kharzeev, Robert
Pisarski and Raju
Venugopalan on the subject of this talk.  Much of the material in the first
three lectures was also presented in the 2003 Moscow Winter School of 
Theoretical Physica and will be published in Surveys in High Energy Physics.

This manuscript has been authorized under Contract No. DE-AC02-98H10886 with
the U. S. Department of Energy.

\end{document}